\documentclass{amsart}
\usepackage{amsaddr}

\RequirePackage{amsthm,amsmath,amsfonts,amssymb, float}

\usepackage{multirow} 
\usepackage{array}
\usepackage{subfig} 
\usepackage{xr} 

\RequirePackage[authoryear]{natbib}
\RequirePackage{graphicx}

\title[Adaptive Block-Based Change-Point Detection]{Adaptive Block-Based Change-Point Detection for Sparse Spatially Clustered Data with Applications in Remote Sensing Imaging}

\author{Alan Moore, Lynna Chu, and Zhengyuan Zhu}
\address{Department of Statistics, Iowa State University
}
\email{alancm@iastate.edu, lchu@iastate.edu, zhuz@iastate.edu}

\begin{document}
\maketitle

\begin{abstract}
We present a non-parametric change-point detection approach to detect potentially sparse changes in a time series of high-dimensional observations or non-Euclidean data objects. We target a change in distribution that occurs in a small, unknown subset of dimensions, where these dimensions may be correlated. Our work is motivated by a remote sensing application, where changes occur in small, spatially clustered regions over time. An adaptive block-based change-point detection framework is proposed that accounts for spatial dependencies across dimensions and leverages these dependencies to boost detection power and improve estimation accuracy. Through simulation studies, we demonstrate that our approach has superior performance in detecting sparse changes in datasets with spatial or local group structures. An application of the proposed method to detect activity, such as new construction, in remote sensing imagery of the Natanz Nuclear facility in Iran is presented to demonstrate the method's efficacy.  

\end{abstract}

\section{Introduction}\label{sec:intro}
Change-point analysis involves the detection and estimation of abrupt changes in a sequence of observations indexed by time or some other meaningful order.  Applications utilizing change-point methods span a wide range of fields; for example disease surveillance, abrupt stock market fluctuations, denial-of-service attacks on networks, or triggering brain activity.  In many of these applications, advancements in data collection technology have resulted in datasets of increasing complexity, heterogeneity, and size. Modern data often involve high-dimensional observations with complex cross-sectional dependencies and non-Euclidean objects, such as images and networks, further compounding the challenge of identifying change-points.

This work is motivated by a remote sensing application where the time series is a sequence of satellite images and the changes of interest are both sparse and spatially correlated. The goal is to identify both time-points and spatial regions where changes occur. In many such time series, observations often exhibit local group structure, such that changes occur in a small, clustered subset of the coordinates or dimensions. These coordinates may possess cross-sectional dependencies or spatial correlations such that coordinates that are close together tend to change together. While some existing change-point methods have been developed to directly handle sparse changes, most do not take into account spatial correlations, and even fewer are directly applicable to non-Euclidean data such as networks. 

To address this gap, we develop a change-point detection approach that is particularly powerful in detecting sparse, spatially clustered changes; we refer to this as the \textbf{A}daptive \textbf{B}lock-based \textbf{C}hange-point \textbf{D}etection (ABCD) method. ABCD is a non-parametric method applicable to high-dimensional and/or non-Euclidean time series. The proposed method imposes no assumptions on the level of sparsity, spatial correlation structures, or the nature of distributional change. To effectively capture sparse, spatially clustered changes over time, we employ an adaptive blocking strategy integrated with a non-parametric testing framework that leverages similarity information among observations.

\subsection{Motivating Application} \label{sec:movation}

\begin{figure}%
    \centering
    \subfloat[\centering December 2019]{{\includegraphics[width=5cm]{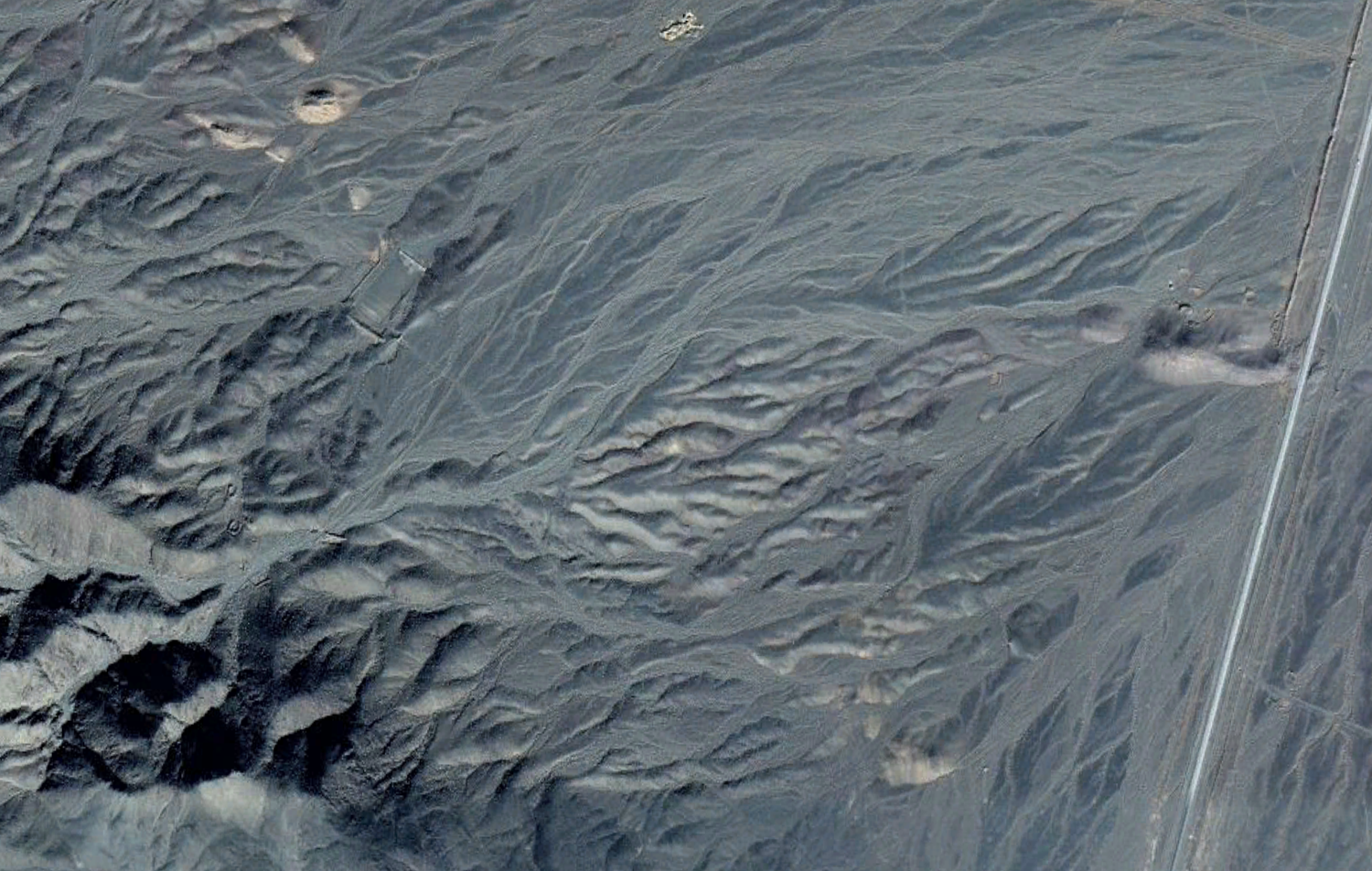}}}%
    \qquad
        \subfloat[\centering December 2020]{{\includegraphics[width=5cm]{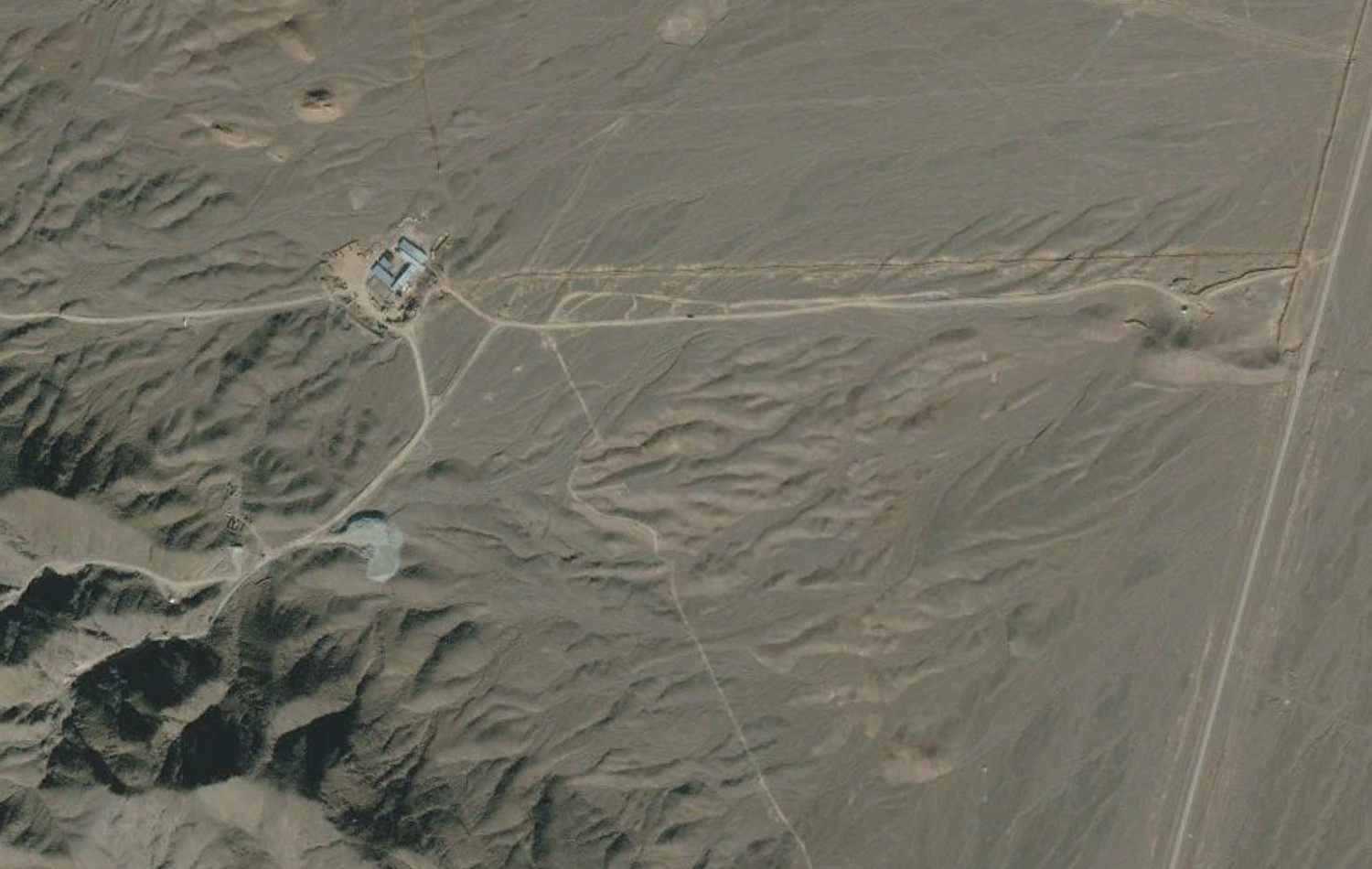}}}%
    \qquad
    \subfloat[\centering January 2022]{{\includegraphics[width=5cm]{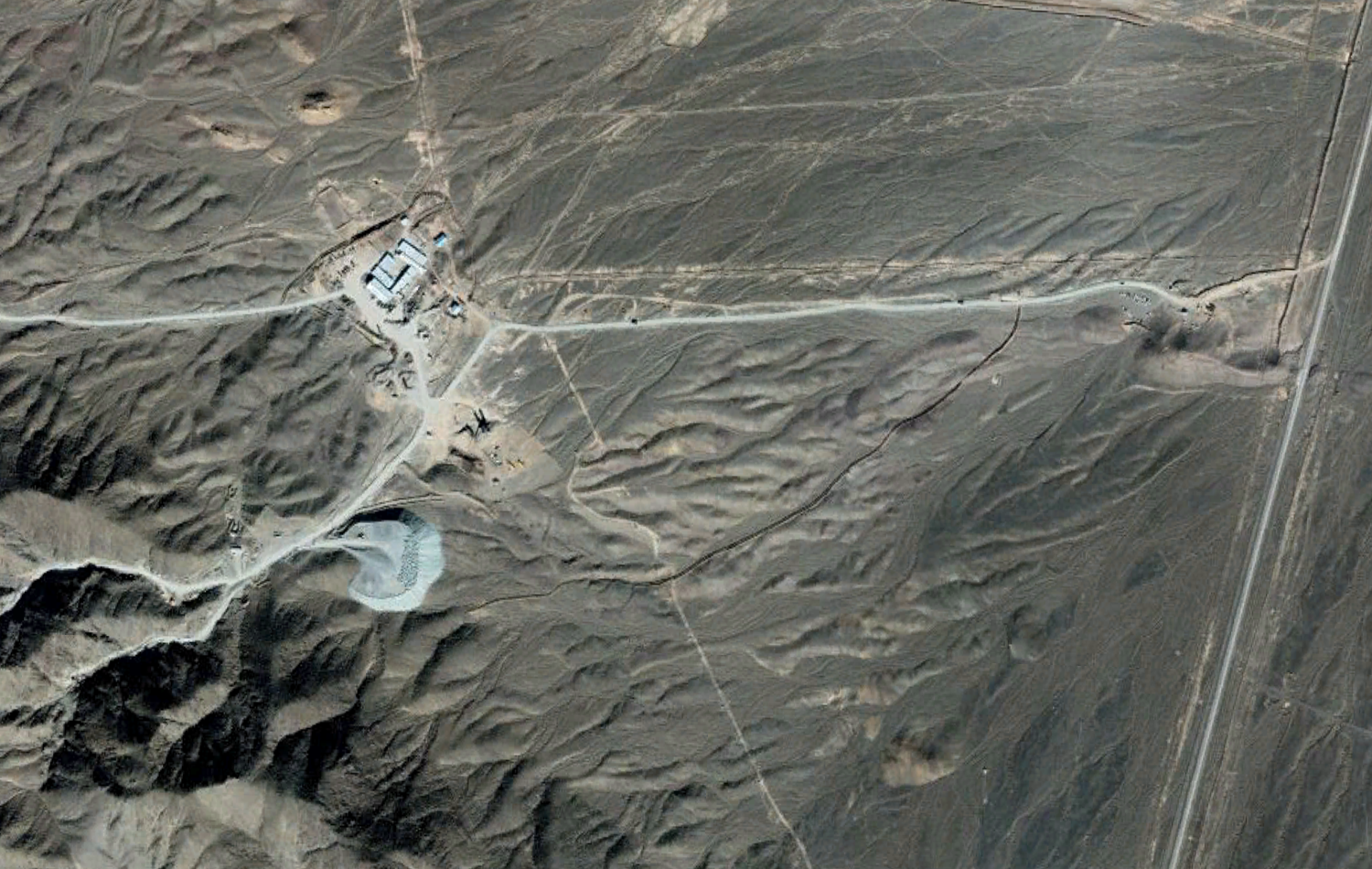} }}%
        \qquad
        \subfloat[\centering September 2023]{{\includegraphics[width=5cm]{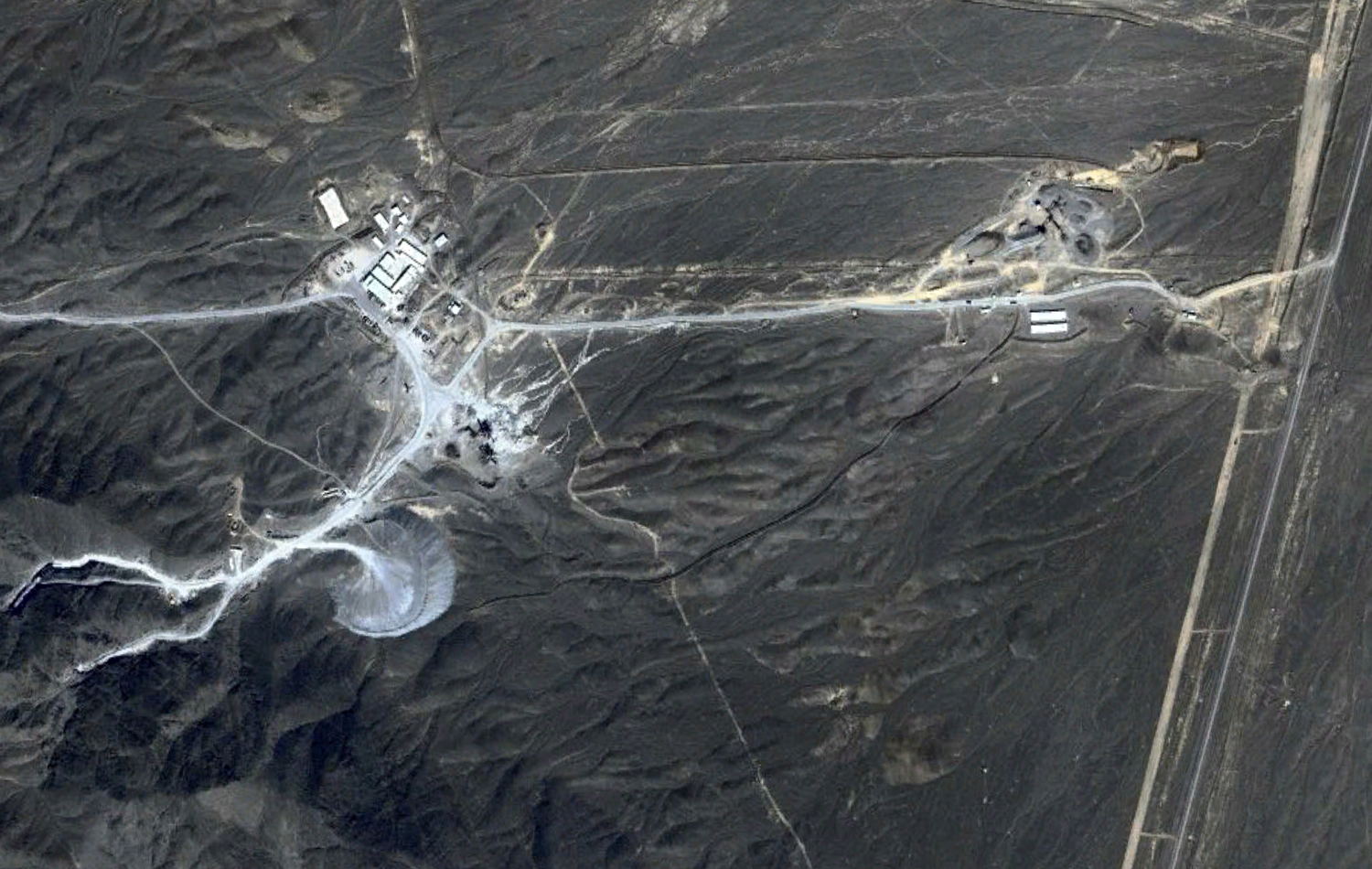} }}%
    \caption{Satellite images of development above Natanz Tunnel Facility, 2019-2023. Maps Data from Google, $\copyright$ CNES, CNES/Airbus and Maxar Technologies.}%
    \label{fig:natanz_pics_over_time}%
\end{figure}

ABCD is motivated by a series of Sentinel-2 satellite images of the Natanz Nuclear Facility in Iran. This is a key site in the country’s nuclear program, where the development of a new underground centrifuge assembly has been ongoing since 2020, as shown in  Figure \ref{fig:natanz_pics_over_time}. The objective is to detect activity, such as new construction, over time. Manual monitoring of this activity via satellite imagery has been valuable to nuclear proliferation experts in determining Iran's progress toward creating nuclear weapons at this site. Ongoing developments would inevitably  have immense implications on the world stage, as exemplified in a recent report by \cite{albright_imagery_2024}. However, the frequency and availability of  sufficiently high-resolution 
reference images for manual monitoring is limited, making it inherently difficult to determine the exact dates of new constructions and activity.  Instead, we utilize Sentinel-2 remote sensing images, which are publicly available via Google Earth Engine. These images have a moderate resolution of 10m per pixel and are taken quite often, with an average of 5 days between Sentinel-2 images of Natanz after cloudy images are removed, offering a more frequent alternative to the high-resolution Earth imagery taken from a variety of sources. 
However, determining when and where new activities occur from the raw Sentinel-2 bands directly is extremely challenging due to the immense size of the dataset, the minuscule size of constructions relative to the overall image domain, and seasonal fluctuations in average band values.

We develop the ABCD method, which is well-suited for this type of problem, as new constructions in images are generally quite sparse, comprising only a small fraction of pixels in the total image. Locations of new activity are often unknown, so no information regarding which coordinates the change will occur is available a-priori. Furthermore, local spatial correlation is present in this data structure; if construction occurs at a given pixel at a point in time, there is a much higher probability that construction will occur in neighboring pixels near the same time. In Section \ref{sec:data_app_section}, we demonstrate that  ABCD is not only powerful for detecting when sparse changes in this sequence of images occur, but is also effective in identifying the specific spatial location of these changes.

\subsection{Existing Literature} \label{sec:lit}

To detect sparse changes in high-dimensional time series, a number of state-of-the-art change-point methods have been proposed. To target sparse mean changes, \cite{jirak_uniform_2015}  proposed an $l_{\infty}$ aggregation across different components, \cite{wang_high_2018} studied a projection-based approach, and \cite{enikeeva_high-dimensional_2019} proposed a scan-statistic testing procedure. Both \cite{wang_high_2018} and \cite{enikeeva_high-dimensional_2019} assume the observations follow a Gaussian distribution.  \cite{liu_unified_2020} and \cite{zhang_adaptive_2022} provide data-adaptive non-parametric methods, attempting to detect both sparse and dense change-points. While effective for changes that occur in a small subset of dimensions, these methods do not directly take into account the cross-sectional dependency between dimensions.  
Methods that address cross-sectional dependency include \cite{cho_change-point_2016}, \cite{bhattacharjee_change_2019}, \cite{bardwell_most_2019} and \cite{hollaway_detection_2024},  but these are not explicitly designed for sparse changes. 

Several change-point methods have recently been proposed for sparse and spatially clustered high-dimensional data. One notable contribution by \cite{cai_estimation_2023} takes into account both sparsity and local group structure, but requires prior grouping information, which may be unrealistic for many data applications. Their approach also assumes the observations are sub-Gaussian and targets detection of mean change only. Another approach from \cite{moradi_hierarchical_2023} targets spatially clustered change-points by first implementing hierarchical clustering of high-dimensional data, followed by a change-point detection algorithm within each cluster. 
\cite{li_2_2024} proposed a MOSUM approach designed to handle dense or spatially clustered signals. However, their approach only targets mean change and becomes computationally expensive to implement as the dimension of the observation increases. 

Another class of methods utilize space-time statistics to detect spatiotemporal clusters; this approach was first introduced for discrete data in \cite{kulldorff_spatial_1997} and later extended to account for spatial correlation by \cite{loh_accounting_2007}, as well as to continuous data applications in space-time using a spatiotemporal Gaussian likelihood ratio scan statistic \citep{kulldorff_scan_2009}.  
However, this method, along with the method from \cite{moradi_hierarchical_2023}, assume spatial stationarity under the null hypothesis - that is, individual dimensions of a given time series are identically distributed (hence, have equivalent mean, variance, and so on).  Otherwise, when this assumption does not hold, appropriate pre-processing of the data may be needed in order to reliably apply these methods. 
Additionally, for large datasets with high-dimensional observations, these tests can become computationally prohibitive.

In the non-Euclidean setting, change-point detection methods applicable to general non-standard data objects are mostly non-parametric and aim to make minimal assumptions about the underlying data distribution. A common strategy is to embed non-Euclidean data objects into a Hilbert or more general metric space, allowing for the application of metric or distance-based approaches.  Recent examples include works by \cite{matteson_nonparametric_2014}, \cite{chu_asymptotic_2019}, \cite{dubey_frechet_2020}, and \cite{jiang_two-sample_2024}. However, none of these methods are specifically designed to target sparse, spatially correlated changes and they may suffer from limited power in such settings.

\subsection{Our contribution}\label{sec:contribution}

In this paper, we develop an adaptive block-based change-point detection (ABCD) approach that is powerful at detecting sparse changes that occur in a small subset of spatially proximal coordinates in a sequence of observations. 
We do not require estimation of any spatial correlation or impose assumptions on the signal's sparsity level.  Crucially, our method is able to detect general distributional changes, whereas the bulk of recent work on sparse or spatially clustered change-points focus exclusively on changes in mean. We build upon the graph-based framework for change-point detection studied in \cite{chu_asymptotic_2019}, in which the authors present a non-parametric framework for change-point detection for generic data types. Their approach implicitly targets dense changes and in the presence of sparse change can suffer from low detection power of change-points (see Section \ref{sec:limitations} for more details). 

Our method uses a strategy of dividing the components of the time series into distinct local blocks, which we denote as a  \textit{blocking structure} for the data. The blocking structure is adaptive to different types of alternatives, allowing our approach to  
capture varying levels of sparsity and spatial cluster sizes.  This enables us to reliably detect changes
without any a-priori knowledge of which subset of coordinates the change occurs in. Motivated by \cite{zhang_spatial_2012}, we propose a test statistic that effectively integrates information across blocks and can boost the signal of smaller, sparse changes.  Moreover, we provide type-I error control via a permutation p-value, which allows user to easily assess the significance and magnitude of the detected change.

The remainder of the paper unfolds as follows. In Section \ref{sec:graph_based_review}, we review the graph-based change-point method studied in \cite{chu_asymptotic_2019}, along with its limitations in detecting sparse, clustered change-points. ABCD is introduced in Section \ref{sec:abcd_single} and illustrative examples are provided for the high-dimensional and image setting. To demonstrate ABCD's performance, simulation studies in Section \ref{sec:sim_studies} compare ABCD to other modern change-point methods under a variety of group sparsity structures.
In Section \ref{sec:data_app_section}, the Natanz facility image data is carefully analyzed under a multiple change-point framework. We highlight ABCD's ability to not only detect changes, but also provide estimates of spatial regions of detected change-points. Concluding remarks are given in Section \ref{sec:Discussion}.

\section{Review of the Graph-based Change-Point Framework} \label{sec:graph_based_review}

\subsection{Graph-based Scan Statistics} \label{sec:graph_stats}

\begin{figure}[h]
\centering
\makebox[\textwidth][c]{\includegraphics[width=1\textwidth]{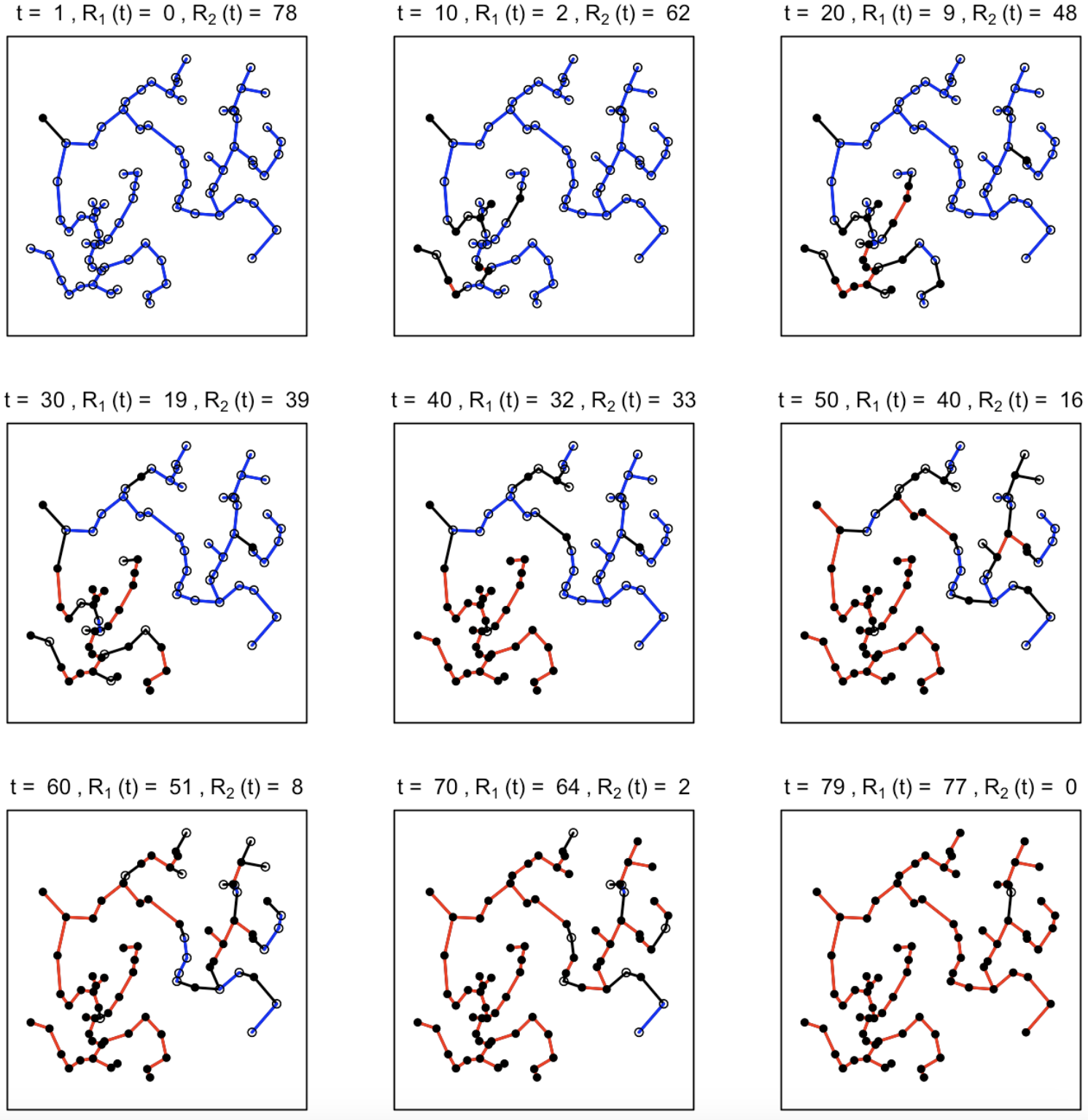}}%
\caption{A depiction of $R_1$ and $R_2$ for different values of $t$. The observations are bivariate, with the first $40$ observations generated from a $\text{Beta}(2, 4)$ distribution, and the remaining $40$ observations  generated from a $\text{Beta}(4,2)$ distribution. The similarity graph, $G$, constructed here is a MST based on Euclidean distance. For a given time $t$, observations are divided into two groups: those before $t$ (shown as hollow circles) and those after $t$ (shown as solid circles).  Edges connecting observations before $t$ are in red; the sum of these edges is $R_1(t)$.  Edges connecting observations after $t$ are in blue; the sum of these edges is $R_2(t)$.}  
\label{fig:graph_based_plots}
\end{figure}

Let $\mathbf{y} = \{y_t$, $t = 1, \hdots, n \}$ be a sequence of observations indexed by time or some other meaningful order. Each $y_t$ is a $d$-dimensional object, such as an image or a network, which may reside in a non-Euclidean space. There possibly exists a time $\tau$ such that $y_t$ follows some unknown distribution for $t \le \tau$, and follows a different unknown distribution for $t > \tau$.  Graph-based change-point analysis provides a general framework for the detection and estimation of such a $\tau$ \citep{chen_graph-based_2023}. The building blocks of the graph-based change-point framework are graph-based two-sample tests; these are a class of non-parametric two-sample tests based on geometric graphs constructed from all observations, such that each observation, $y_t$, corresponds to a node in the graph. These omnibus two-sample tests are designed to assess whether or not the two samples 
are from the same distribution. The graph is usually derived from a distance or a generalized dissimilarity on the sample space such that observations that are more similar in some sense are more likely to have an edge connecting them. Examples of similarity graphs include a minimum spanning tree (MST) \citep{friedman_multivariate_1979}, which is a tree connecting all the observations in such a way that the total distance across edges is minimized, or  a $k$-MST, which is the union of a MST and $k-1$ spanning trees, where the $i^{th}$ $(i > 1)$ spanning tree does not contain any edges from the first $i-1$ spanning trees. Other examples include the $k$-nearest neighbor graph ($k$-NNG), where each observation has an edge connecting to its $k$ nearest neighbors \citep{henze_multivariate_1988}, or the graph may be be user-specified and constructed from domain knowledge. These graph-based two-sample tests have been shown to be universally consistent and computationally efficient, making them useful tools for a broad range of modern applications \citep{bhattacharya_asymptotic_2020}. 

In the graph-based change-point framework, the graph-based two-sample tests are adapted to the scan statistic setting. Consider that each time $t$ divides the sequence into two samples: those observations that occur before $t$ and those observations that occur after $t$. Then, for each time $t$, the graph-based test determines whether there is a difference in distribution. We use $G$ to denote both the similarity graph constructed on $\mathbf{y}$ and its set of edges when its vertex set is implicitly obvious. Let $g_i(t) = I\{i>t\}$ denote an indicator function equal to 1 only if time $i$ is after time $t$.  

There are two quantities of interest we calculate from the similarity graph, $G$: 
\begin{align*}
R_1(t) & = \sum_{(i,j)\in G} I\{g_i(t) = g_j(t) = 0\}, 
& R_2(t) & = \sum_{(i,j)\in G} I\{g_i(t) = g_j(t) = 1\}.
\end{align*}

\noindent These are referred to as the within-sample edge counts. Specifically, $R_1(t)$ is the number of edges connecting observations prior to time $t$ (sample 1) and $R_2(t)$ is the number of edges that connect observations after time $t$ (sample 2). In Figure \ref{fig:graph_based_plots}, we illustrate the construction of the graph and computation of $R_1(t)$ and $R_2(t)$ for a time series of bivariate observations with Beta marginals, with $n = 80$ and a distributional change at time $t = 40$.  

In \cite{chu_asymptotic_2019}, (standardized) combinations of the within-sample edge-counts were used to construct scan statistics. In particular, we focus on the \textit{max-type edge-count statistic}:
\begin{align*}
M(t) & = \max(Z_w(t),| Z_\text{diff}(t) |),
\end{align*}
\noindent where 
\begin{align*} 
Z_w(t) &= \frac{R_w(t) - E(R_w(t))}{\text{var}(R_w(t))}, \thickspace \\
Z_\text{diff}(t) & = \frac{R_\text{diff}(t) - E(R_\text{diff}(t))}{\text{var}(R_\text{diff}(t))},\\
R_w(t) & = q(t) R_1(t) + p(t)R_2(t), \hspace{3mm} p(t) = \frac{(t-1)}{(n-2)},\hspace{3mm} q(t) = 1 - p(t), \\
R_\text{diff}(t) & = R_1(t) - R_2(t). \\
\end{align*}

\noindent The quantities $Z_\text{diff}(t)$ and $Z_w(t)$ each have their own niche. While $Z_w(t)$ is designed to detect mean changes, $Z_\text{diff}(t)$ is intended to detect changes in variance. 
Since the max-type statistic $M(t)$ is a function of $Z_w(t)$ and $Z_\text{diff}(t)$,
the combination of these two statistics allow us to detect general distributional changes. Moreover, the construction of $M(t)$ is able overcome the curse of dimensionality, making it particularly effective when the dimension of the observation is moderate to high \citep{chu_asymptotic_2019}. Simulation studies presented in \cite{chen_graph-based_2023}, and references therein, demonstrate that the approach is powerful even for changes beyond mean and variance.

Using combinatorics, analytical formulas for $E(R_w(t))$, $E(R_{\text{diff}}(t))$, $\text{var}(R_w(t))$ and $\text{var}(R_{\text{diff}}(t))$ can be computed under the permutation null distribution, where $1/n!$ probability is placed on each of the $n!$ permutations of $y_t$, $t = 1, \hdots, n$. The analytical expressions are provided explicitly in \cite{chu_asymptotic_2019} and are omitted here for ease of presentation. 

The corresponding max-type scan statistic is
$$\max\limits_{n_0 \leq t \leq n_1} M(t),$$

\noindent where $n_0$ and $n_1$ are pre-specified constraints for the range of the change-point $\tau$. The null hypothesis of no change-point is rejected when the maximum of the scan statistic is larger than a pre-specified threshold. 

\subsection{Limitations of the existing approach under sparse, spatially clustered alternatives} \label{sec:limitations}

In the existing graph-based framework, the similarity graph is constructed using all pairwise inter-point distances. Without prior knowledge regarding the alternative, a standard choice to construct the graph may be $L_1$ or $L_2$ distance. Naturally, if the strength of signal is relatively sparse and only occurs in a subset of spatially clustered dimensions, the signal can be diluted under standard distance metrics and the existing graph-based approach can have limited power in detecting the change-point. To illustrate this, consider a simple simulation setting where we have a sequence of $n = 500$ observations each of dimension $d = 1000$. A change happens in the middle of the sequence ($\tau = 250$) and the change is in mean and variance ($\mathcal{N}(\textbf{0},I_d)$ versus $\mathcal{N}(\mu,\sigma^2 I_d)$ where $\| \mu \| = 1$ and $\sigma^2 = 1.05^2$). 

Across all settings, we keep the $L_2$ distance from $\textbf{0}$ to be the same ($\| \mu \| = 1$) after $\tau$, but we control the portion of dimensions that undergo a change, such that only the first $D$ dimensions change and hence the change is clustered based on the ordering of components. We then report the power of the max-type edge-count statistic $M$ for the existing graph-based method (denoted as gSeg), and compare it to the power of the proposed ABCD method. Power is estimated as the number of trials (out of 100) that detect a significant change-point.  Significance at $\alpha = 0.05$ is determined using a permutation p-value with $B=1000$ permutations. Both ABCD and gSeg are constructed from 50-MSTs.

 \begin{table}[H]  \centering
\caption{Number of trials out of 100 where the change-point method can reject the null hypothesis of no change-point at significance level  $\alpha = 0.05$.  Multivariate Gaussian data is generated such that before the change, data is drawn from $\mathcal{N}(0,I_d)$ and after the change data is drawn from $\mathcal{N}(\mu,\sigma^2 I_d)$, with $\| \mu \| = 1$ and $\sigma^2 = 1.05^2$. The length of the sequence is $n = 500$ and the dimension of each observation is $d = 1000$. Here $D$ controls the sparsity, with the mean change occurring only in the first $D$ dimensions of observations after $\tau = 250$. }
\begin{tabular}{ |c|cccccc| } 
\hline
$M$ & $D = 1000$ & $D = 500$ &  $D = 200$ &  $D = 100$  &  $D = 50$ &  $D = 10$ \\ 
\hline
gSeg Power & 100 & 100 & 91 & 42 & 27 & 21\\
\hline
ABCD Power  & 100 & 100 & 100 & 100 & 99 & 100\\
\hline
\end{tabular}
\label{table:motivate}
\end{table}

We observe that when the signal is spatially clustered and the change becomes increasingly sparse, gSeg's power diminishes, while ABCD is able to preserve its power. This dwindling of detection power in gSeg motivates a change-point method that can account for potentially sparse changes by leveraging the spatial proximity of dimensions. Even when the change is relatively dense ($D=1000$ and $D=500$), the adaptive nature of ABCD allows it to perform just as well as gSeg.  

\section{ABCD Method for the Single Change-point Case} \label{sec:abcd_single}

\subsection{Problem Formulation} \label{sec:problem_form}
Consider a sequence of observations $\textbf{y} = \{y_t: t = 1, \hdots, n\}$, where $y_t$ may be a high-dimensional vector ($d>n$) or a non-Euclidean data object.

We are interested in testing the null hypothesis
$$H_0: y_t\sim F_0,\thickspace \thickspace t = 1, \hdots, n $$
against the single change-point alternative
\begin{align*} 
H_a: \exists \thickspace 1 \le \tau < n \text{ such that } y_t \sim
\begin{cases} 
& \negthickspace \negthickspace F_0,\thickspace \thickspace t = 1, \hdots, \tau, \\
& \negthickspace \negthickspace F_1, \thickspace \thickspace t = \tau+1, \hdots, n 
\end{cases} 
\end{align*}

\noindent where $F_0$ and $F_1$ are two distinct unspecified probability measures.

In what follows, we focus on the single change-point alternative. We discuss methods for extending the methodology to deal with multiple change-points in Section \ref{sec:data_app_section}. Under the alternative, we define a \textit{change region} $\mathcal{C}$ to be a spatially contiguous subset of dimensions of $\textbf{y}$ such that some of the dimensions in $\mathcal{C}$ undergo a change in distribution. 
Not all dimensions within a change region necessarily change, nor does the change have to happen in consecutive dimensions.  Importantly, we assume no knowledge about the size, shape, or location of $\mathcal{C}$. 
This is often the case for remote sensing or pandemic applications, the latter of which was previously investigated in the spatial scan statistic literature \citep{kulldorff_spatial_1997}. 
Later in Section $\ref{sec:sim_studies}$, we study scenarios where the strength of spatial correlation in the change region varies.

\subsection{Proposed Method} \label{sec:method_proposal}

Suppose a change occurs in a potentially sparse change region $\mathcal{C}$ that exhibits a local dependency structure. Our proposed method ABCD attempts to exploit this local alternative structure by dividing the dimensions of the time series into contiguous blocks. The strength of signal over time is summarized in each block by a graph-based scan statistic. If a change region falls mostly within a specific block, the signal of the change-point will be greatly boosted. While the blocks may not exactly capture the specific size or shape of a change region, which may be split across multiple blocks, we overcome this issue by constructing an ensemble test statistic using information from multiple \textit{blocking structures}.  

For a given blocking structure, we opt to divide dimensions into non-overlapping blocks of roughly equal size, rather than using an exhaustive test to analyze all possible (overlapping) blocks of a given size, as is done in the space-time scan statistic of \cite{kulldorff_scan_2009}. This helps mitigate high computational complexity and since our approach tests over multiple blocking structures, ABCD is well-equipped to capture the signal of a change region of unknown size, shape and location within the time series. 

We describe notation and details for blocking structures in Section \ref{sec:blocking_structures}, while the details of how ABCD integrates information across blocks to generate an ensemble test statistic is discussed in Section \ref{sec:alg_details}. Examples for high-dimensional and image data provided in Sections \ref{sec:hd_example} and \ref{sec:2d_ABCD_desc}, respectively. 

\subsubsection{Blocking Structures} \label{sec:blocking_structures}

To develop methodology that can efficiently leverage spatial information, we assume that each component of $\textbf{y}$ has associated coordinates in some metric space which are ordered and equidistant from their neighbors. 
If $y_t$ is non-Euclidean, we assume that it can be embedded within $\mathbb{R}^q$ space with distance metric $\nu(\cdot)$, together forming metric space $(\mathbb{R}^q, \nu)$. An example of such a data structure would include a grid of points in $\mathbb{R}^2$ or a cube of voxels in $\mathbb{R}^3$. The components of $y_t$ can be decomposed as $d = d_1 \times d_2 \times\ldots \times d_q$, such that along the $r^{th}$ dimension in the metric space where $r = 1,\ldots,q$, we can divide $y_t$ into $d_r$ array slices with $d_1 \times \ldots \times d_{r-1} \times 1 \times d_{r+1} \times \ldots \times d_q$ components. With this decomposition, we can label the components of the time series more precisely as $(h_{1,\ldots, 1}, h_{1,\ldots, 2}, \ldots, h_{d_1, \ldots, d_q})$. In the high-dimensional case, with $q = 1$ and $d = d_1$, the notation greatly simplifies such that $y_t$ has components labeled $(h_1, \ldots, h_d)$. In the case of an image time series, then $d = d_1 \times d_2$, where $d_1$ represents the number of rows of pixels, and $d_2$ represents the number of columns of pixels, with component labels $(h_{1,1}, \ldots, h_{d_1,d_2})$. 




When $q = 1$, for a single blocking structure, we divide the $d$ components of $\textbf{y}$ into $P$ blocks $B_1,  \ldots, B_{P}$ of relatively equivalent size, with $B_1, \ldots, B_{P-1}$ each containing $\lfloor \frac{d}{P} \rfloor$ components, and a final remainder block $B_{P}$ containing the remainder $d - (P-1)(\lfloor \frac{d}{P} \rfloor)$ components. Hence $B_1 = (h_1, ..., h_{\lfloor \frac{d}{P} \rfloor})$, $B_2 = (h_{\lfloor \frac{d}{P} \rfloor + 1}, ..., h_{2\lfloor \frac{d}{P} \rfloor})$, and so on. 
To extend this to multiple blocking structures, suppose that we have $\mathcal{S}$ distinct blocking structures. We define the vector $\textbf{P} = (P_1, ..., P_\mathcal{S})^T$, such that $P_s \in \mathbb{N}$ represents the number of blocks for associated blocking structure $s$ and its corresponding blocks $B_j^{s}$, $j = 1 \hdots, P_s$. Without prior knowledge on the size of the change region, a range of values for $\textbf{P}$ should be considered, with smaller numbers of blocks corresponding to larger block sizes that can detect changes in larger clusters, and larger numbers of blocks leading to smaller block sizes that are more effective in detecting changes in smaller clusters. The more blocking structures included in the overall procedure, the fewer blind spots there will be in detecting an unknown change region, with the trade-off of increased computational complexity. On the other hand, if one \textit{does} have prior domain knowledge about the size, shape or location of the change region, then it may be sufficient to choose one blocking structure to target this type of alternative. 

For general $q $, to construct a single blocking structure (indexed by $s$), one may create blocks by dividing observations $\textbf{y}$ along each of the $q$ dimensions of the metric space. As in the case of $q = 1$, a total of $\mathcal{S}$ such blocking structures can be created 
to adaptively capture different types of alternatives. For the remainder of the paper, we focus on cases where $q \le 2$, with additional clarification for the case of $q= 2$ given in Section \ref{sec:2d_ABCD_desc}. 

\begin{figure}[t]
\centering
\makebox[\textwidth][c]{\includegraphics[width=0.95\textwidth]{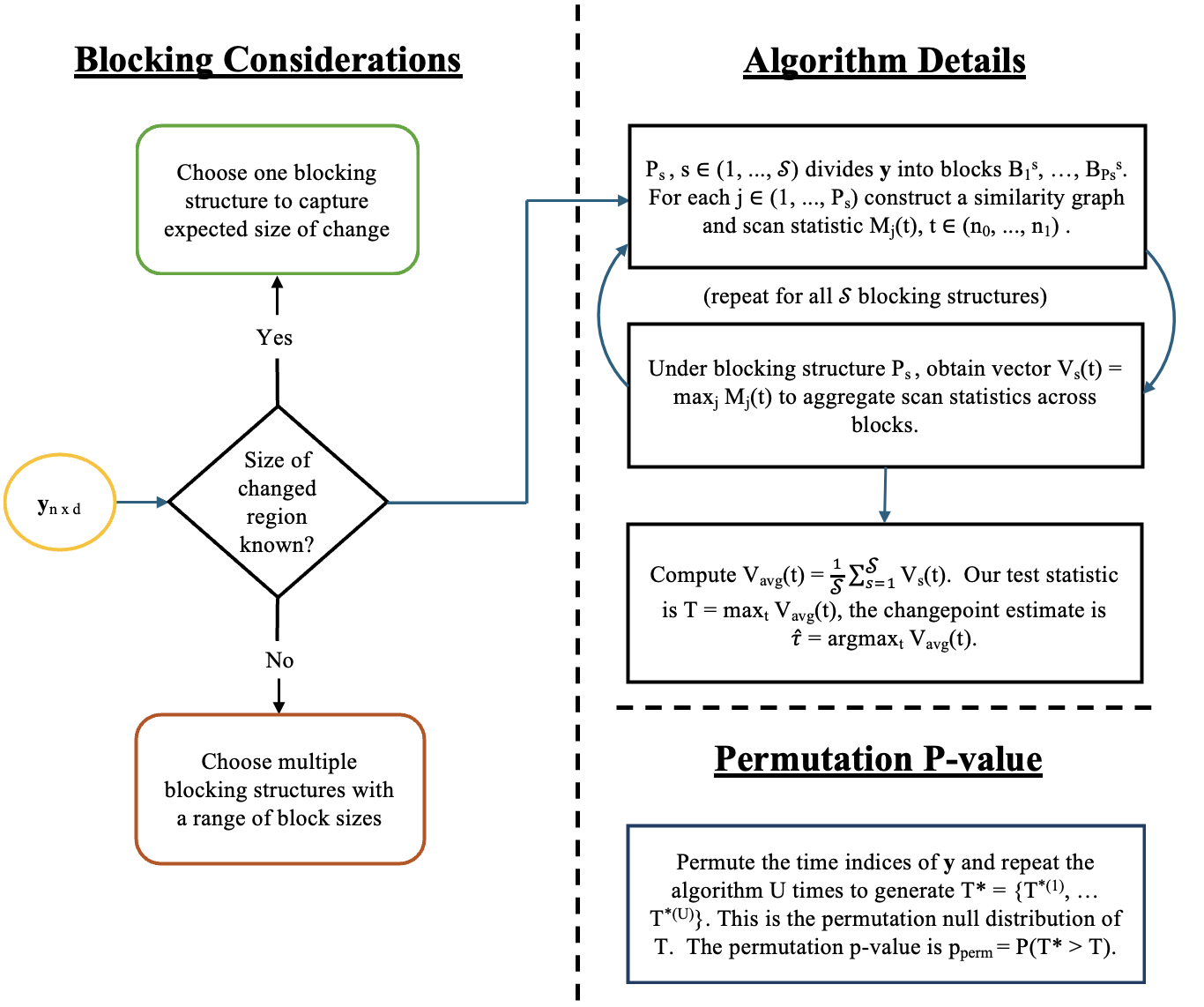}}%
\caption{ABCD method flowchart. For ease of notation, here $
y_t$ is a $d$-dimensional observation. For general settings, the same workflow applies, with more careful treatment of the blocking indices.}
\label{fig:flowchart}
\end{figure}

\subsubsection{Details of ABCD algorithm} \label{sec:alg_details}

A high-level overview of our proposed workflow can be found in  Figure \ref{fig:flowchart}. For ease of presentation, suppose $\textbf{y}$ is a $d$-dimensional vector. Then for a given value $P_s \in \textbf{P}$, the time series $\textbf{y}$ is divided up into blocks $B^s_1,  \ldots, B^s_{P_s}$ indexed from $j = 1, \ldots, P_s$. We treat each block $B^s_j$ as a distinct multivariate time series. 
For each block, we construct a similarity graph from all $n$ observations in the sequence. A discussion on the choice of similarity graph can be found in Section A2 of the Supplement. Given a block's similarity graph, a graph-based scan statistic  is computed 
that summarizes the strength of signal across time, resulting in a total of $P_s$ scan statistics, one for each block. We let $M_j(t)$ represent the max-type graph-based scan statistic generated for block $B_j$ at time $t$.  To aggregate the scan statistics across blocks, we compute $$V_s(t) = \max\limits_{1 \le j \le P_s} M_j(t),$$

\noindent a vector consisting of the maximum value across the $P_s$ scan statistics found for time-point $n_0 \leq t \leq n_1$.  Observations near the boundary often exhibit larger volatility, which can result in unstable change-point conclusions. To protect against this phenomenon, it is common practice to exclude boundary observations as candidates for change-points. To make things simple, we let $n_1 = n - n_0$ and unless other specified $n_0 = \lfloor{0.05n} \rfloor$. Note that if $q > 1$, to compute $V_s(t)$ we would take a maximum at each time $t$ across $P_{s1}\times \ldots \times P_{sq}$ blocks.

We repeat this procedure for all $P_s \in \textbf{P}$ to obtain $\mathcal{S}$ vectors $V_1(t), V_2(t), ..., V_\mathcal{S}(t)$.  Then, to test $H_0$ versus $H_1$, the average of these maximum vectors at each time-point is used:
$$V_{avg}(t) = \frac{1}{\mathcal{S}}\sum_{s=1}^\mathcal{S} V_s(t).$$

\noindent The estimated change-point $\hat{\tau}$ is 
$$\hat{\tau} = \text{arg}\max\limits_{t} V_{avg}(t).$$

\noindent The motivation behind this final averaging step is that, while certain blocking structures are better suited for detecting change-points in specific change regions, even sub-optimal blocking structures may still capture useful signals if a change is present. By averaging across multiple blocking structures, we harness the strengths of each, ensuring that the method is able to detect a wide range of levels of sparsity and sizes of change region.  This provides stability in our approach by ensuring that information, even from the sub-optimal blocking structures, is not neglected. This should also give practitioners more leniency in the choice of blocking structures, as choosing a single best blocking structure is less critical.

The null hypothesis of no change is rejected if the maximum of $V_{avg}(t)$ is larger than a threshold. Let $T = \max\limits_{t} V_{avg}(t)$. To assess the statistical significance of $T$, a permutation-based procedure is used to generate a $p$-value for $T$. Since we do not assume independence between blocks or any specific spatial structure, the underlying null distribution of $T$ is highly complex. Permutation resampling allows us to approximate the null distribution without imposing strict assumptions. The permutation distribution places 1/n! probability on each of the n! permutations of $\mathbf{y}$. In practice, the permutation null distribution is approximated by the empirical distribution of $T$ for a large number $U$ of permutations. This resampling process allows us to obtain $T^\star = (T_*^{(1)}, T_*^{(2)}, ..., T_*^{(U)})$. The permutation p-value is calculated as $p_{perm} = P(T^*>T)$.

\subsection{Example for high-dimensional data}\label{sec:hd_example}

To provide practical insight on how the procedure works, we give an example for the high-dimensional case. Let $\textbf{y}$ be a time series with $n = 100$ generated from a multivariate standard Gaussian distribution with $d = 500$ dimensions labeled $(h_1, ..., h_{500})$. For $t >50$, the first $50$ dimensions $(h_1, ..., h_{50})$ undergo a mean change such that the marginal mean is $\mu = 0.25$. 
Assuming no knowledge about the alternative structure, we choose $\textbf{P} = (P_1 = 1, P_2 = 4, P_3 = 10, P_4 = 20)^T$ to cover a broad spectrum of potential change regions.

\begin{figure}[t]
\centering
\includegraphics[scale = 0.5]{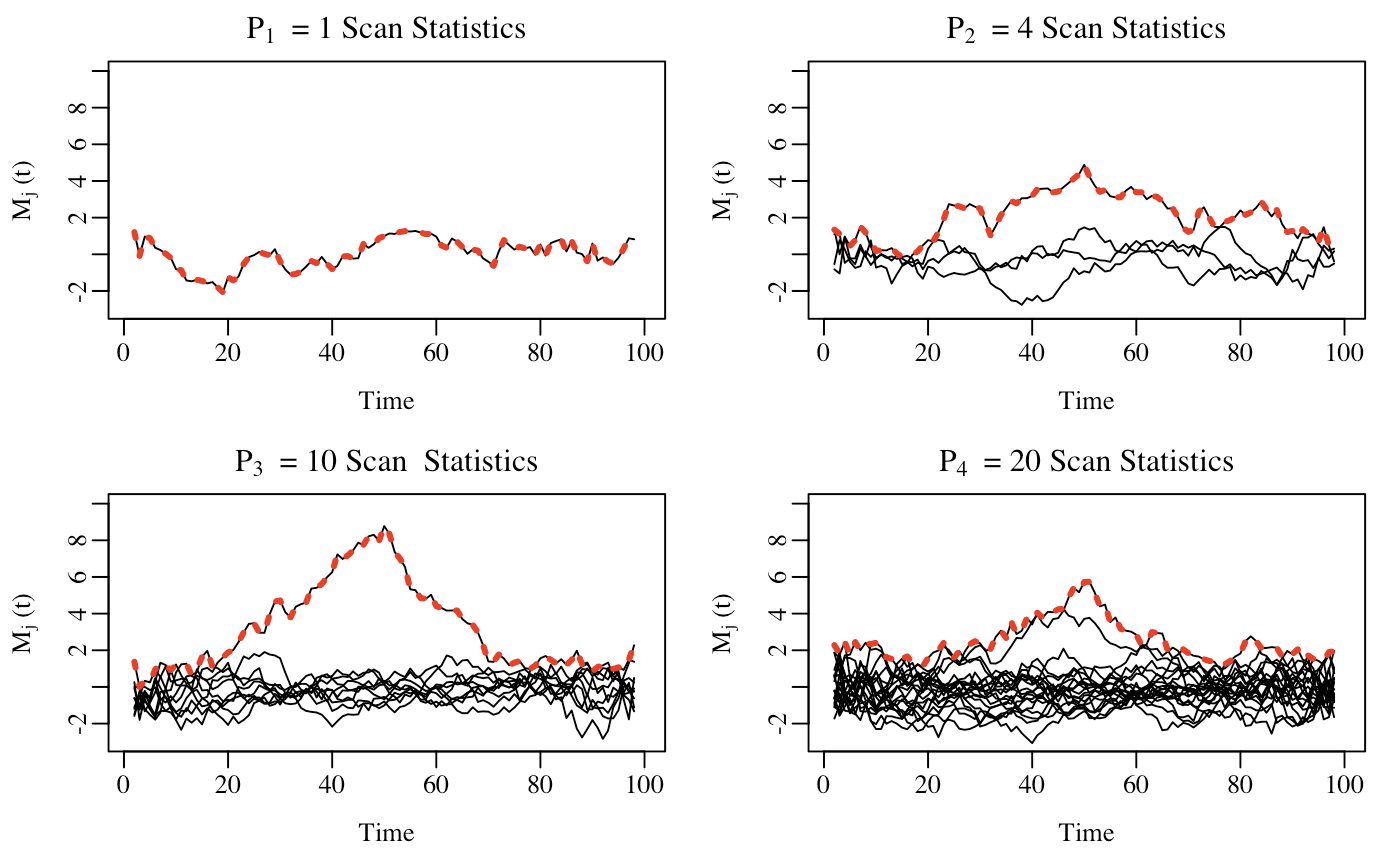}
\caption{Plots of scan statistics $M_j(t)$ are shown in black for different blocking structures. Here $\textbf{P} = (1, 4, 10, 20)^T$. The bold red line indicates the maximum vector $V_s(t)$ for structures $s = 1, 2, 3, 4$.}
\label{fig:block_scans_fig}
\end{figure}

\begin{figure}[h]
\centering
\includegraphics[scale = 0.4]{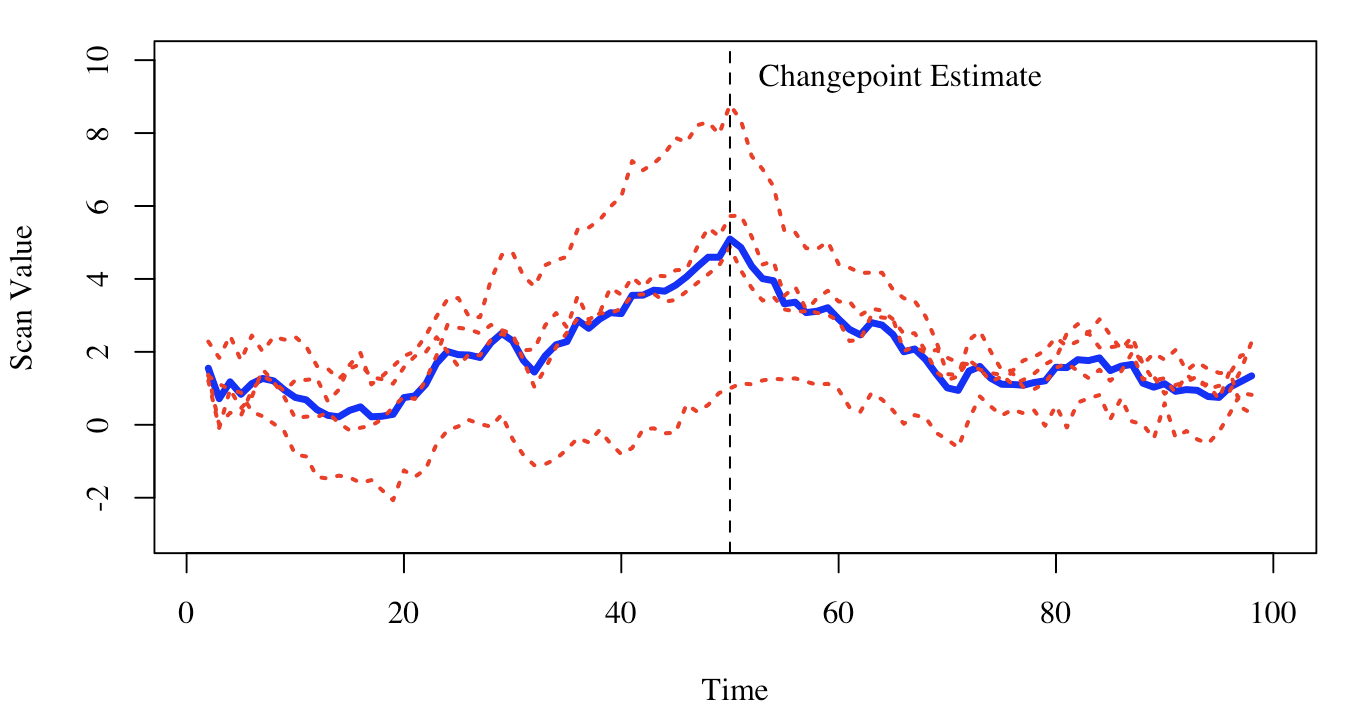}
\caption{Plot of maximum vectors $V_1(t), V_2(t), V_3(t), V_4(t)$ associated with each blocking structure, with their vector average $V_{avg}(t)$ in blue.}
\label{fig:max_vecs_fig}
\end{figure}

Then, for each block within the blocking structure, we construct a $k$-MST  and obtain the graph-based scan statistic $M_j(t)$, $j = 1, \hdots, P_s$. 
For each blocking structure $s \in (1, 2, 3, 4)$, a vector $V_s(t)$ of the maximum scan value across blocks is obtained for $n_0 \le t \le n_1$. This is illustrated for each blocking structure specified by $\textbf{P}$ in Figure \ref{fig:block_scans_fig}, which displays plots of $M_j(t)$ for each structure in black, along with the maximum value across blocks over time, $V_s(t)$, in red. The blocking structure corresponding to $P_3 = 10$ blocks has the scan statistic with the largest peak at the true change-point; this is expected given our setting since this blocking structure most precisely isolates the change region within a single block. Finally, the algorithm takes the average $\frac{1}{4} \sum_{s=1}^4 V_s(t)$ to get a final vector of scan statistics $V_{avg}(t)$. We estimate the location of the change-point to be $\hat{\tau} = 50$, as shown in Figure \ref{fig:max_vecs_fig}.

\begin{figure}[h]
\centering
\makebox[\textwidth][c]{\includegraphics[width=1\textwidth]{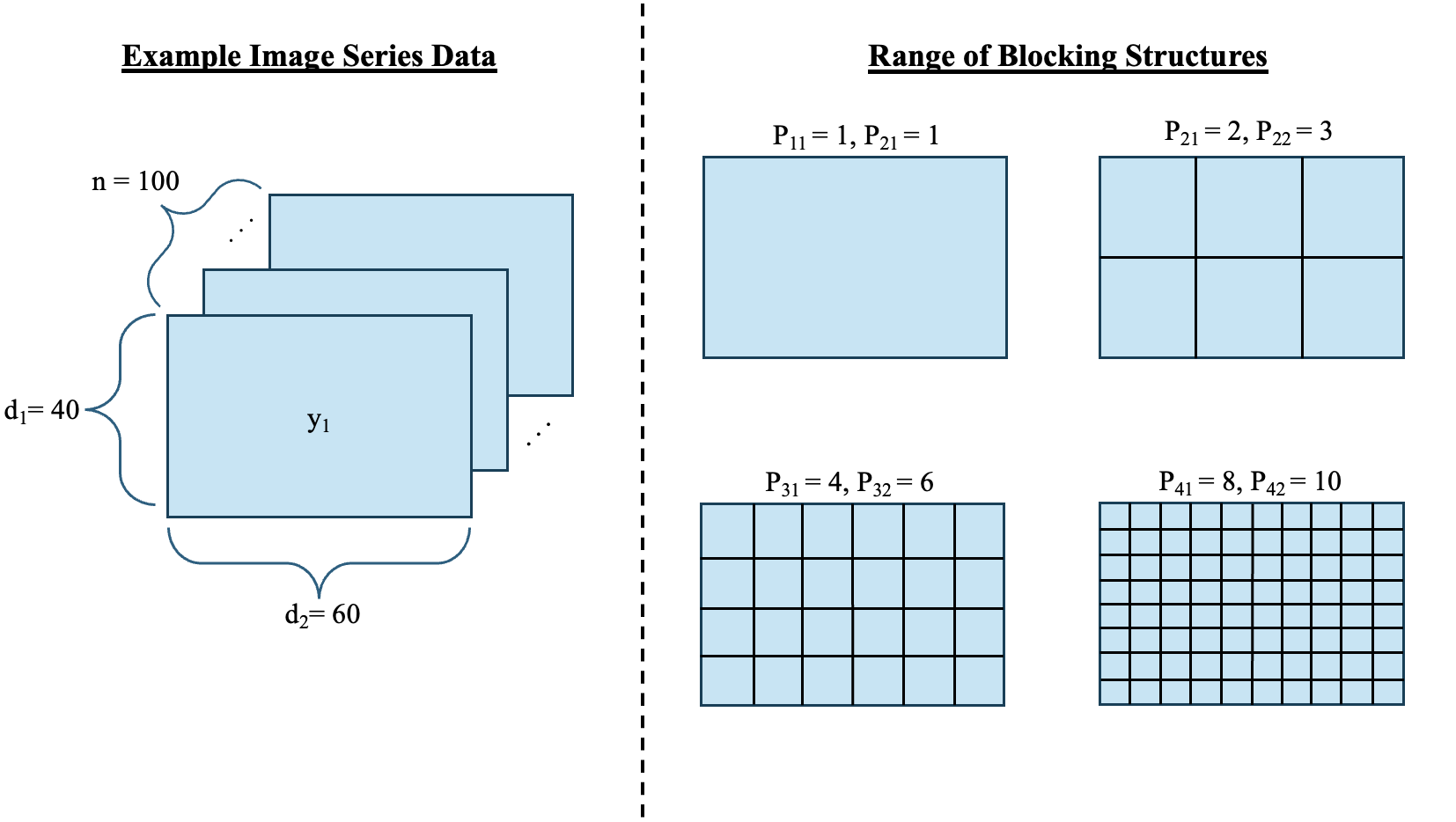}}%
\caption{Visual representation of 4 blocking structures for a time series of $n = 100$ images with $40 \times 60$ pixels. The block sizes in each structure range from large to small to capture different sizes of change region.}
\label{fig:2d_example}
\end{figure}

\subsection{Blocking for a series of images}\label{sec:2d_ABCD_desc}

Consider a time series $\textbf{y}$ consisting of $n$ images with $d_1$ rows and $d_2$ columns of pixels (components). For a single blocking structure indexed by $s$, we divide the image into $P_{s1}$ rows and $P_{s2}$ columns, resulting in a total of $P_{s1} \times P_{s2}$ blocks. The $d_1$ rows of pixels are divided into $P_{s1}$ block rows, with the first $P_{s1} - 1$ block rows having height $\lfloor \frac{d_1}{P_{s1}} \rfloor$, and a final remainder row of blocks  having a height of $d_1  - (P_{s1} - 1)(\lfloor \frac{d_1}{P_{s1}} \rfloor)$ pixels. The columns can be blocked in a similar manner.  
When using $\mathcal{S}$ total block structures, there will be $\mathcal{S}$ pairs representing the number of rows and columns of blocks.  Let $\textbf{P} = [\textbf{P}_1, \textbf{P}_2]$ be a $\mathcal{S} \times 2$ matrix where
$\textbf{P}_1 = (P_{11}, ..., P_{\mathcal{S}1})^T$ and $\textbf{P}_2 = (P_{12}, ..., P_{\mathcal{S}2})^T$  represent the number of rows of blocks and number of columns of blocks in each blocking structure, respectively.  

To give a specific example for the image setting, consider a time series of $n = 100$ images of $40 \times 60$ pixels.  Without a-priori knowledge about the change region, we set $\mathcal{S} = 4$ distinct blocking structures given by 

$$\textbf{P} =  
  \left[ {\begin{array}{cccc}
   1 & 2 & 4 & 8 \\
   1 & 3 & 6 &10\\
  \end{array} } \right]^T.
$$

\noindent Hence, the first blocking structure contains a single block, the second contains $6$ blocks, the third contains $24$ blocks and the final contains $80$ total blocks. Figure \ref{fig:2d_example} illustrates how the pixels will be separated into $4$ distinct blocking structures. 
Apart from the practical differences in blocking in the image setting, the same workflow as laid out in Section \ref{sec:alg_details} is implemented to construct an ensemble test statistic.

\section{Simulation Studies} \label{sec:sim_studies}

We present simulations studies to assess ABCD's ability to detect change-points which are spatially clustered to varying degrees. In Section \ref{sec:sim_studies1}, we examine ABCD's accuracy and power in detecting sparse changes in the high-dimensional setting; we compare our method to Inspect \citep{wang_high_2018}, Double CUSUM (DC)  \citep{cho_change-point_2016}, and the scan statistic SCAN \citep{enikeeva_high-dimensional_2019}. In Section \ref{sec:sim_studies2}, we evaluate ABCD's performance in the image setting. We compare ABCD to 2WayMOSUM \citep{li_2_2024} and the space-time scan statistic (SSS) \citep{kulldorff_scan_2009},  which incorporate spatial information into their respective procedures and provide locational results for change-points.

\subsection{High-Dimensional Simulations}
\label{sec:sim_studies1}

The performance of the ABCD method is compared to other modern change-point detection methods in scenarios of varying spatial concentration and sparsity. Our focus is primarily on high-dimensional Gaussian data sequences, since this setting is largely beneficial for the competing methods.  We generate $n=200$ observations from a $d=1000$ dimensional multivariate Gaussian distribution; observations after time $t = 120$ undergo a change in distribution. 

We consider three simulation scenarios: mean change, variance change, and mean change when the dimensions are spatially correlated under $H_0$. 
For each simulation scenario, we consider a level of sparsity denoted by $D$, such that $D = 50$ dimensions experience a change after $\tau=120$. Additional experiments using $D = 25$ and $D = 100$ can be found in Section A3 of the Supplement. For each level of sparsity, the change occurs in a change region $\mathcal{C} \thinspace \subseteq \thinspace (h_1, \ldots, h_{1000})$ such that only a specified proportion  (denoted as $p_{\mathcal{C}}$) of dimensions  within $\mathcal{C}$ are randomly selected to change after $\tau = 120$. 
We consider three proportions: $p_\mathcal{C} = \frac{1}{4}, \frac{1}{2}$, and $1$. Let $H_D \subseteq \mathcal{C}$ represent the index set of dimensions undergoing a change, where $|H_D | = D$.  Then, for given values of $D$ and $p_{\mathcal{C}}$, the change region is defined to be $\mathcal{C} = (h_1, h_2, \ldots, h_{D  /p_{\mathcal{C}}} )$. Observe that the change region shrinks as $p_\mathcal{C}$ increases, thus the expected density of the change occurring within $\mathcal{C}$ varies depending on the value $p_{\mathcal{C}}$. 
The more densely packed the change within $\mathcal{C}$ is, the more we expect ABCD will be able to leverage the spatial proximity of the changed dimensions to boost the change-point signal, with the extreme case being $p_{\mathcal{C}} = 1$ when $H_D = \mathcal{C}$. However, this spatial compactness may lead to stronger spatial correlation between changed components due to their increased proximity, potentially hindering change-point detection. 

In the first scenario, we consider multivariate Gaussian data with mean change. Explicitly,  we set

 $$ y_t \sim
\begin{cases} 
& \negthickspace \negthickspace \mathcal{N}(\textbf{0}, I_{1000}),\thickspace \thickspace t = 1, \hdots, 120, \\
& \negthickspace \negthickspace \mathcal{N}(\mu \textbf{1}_{H_D}, I_{1000}), \thickspace \thickspace t = 121, \hdots, 200
\end{cases} $$

\noindent where $\textbf{1}_{H_D}$ is a $1000 \times 1$ vector whose $k^{th}$ entry is $1$ if $h_k \in H_D$ and $0$ otherwise, and $\mu = \sqrt{\frac{2.25}{D}}$. 
Hence, the total magnitude of mean change in $L_2$ is $\| \mu \textbf{1}_{H_D} \| = 2.25$.  


In the second scenario, we generate multivariate Gaussian data with variance change. 
Here, we set

 $$ y_t \sim
\begin{cases} 
& \negthickspace \negthickspace \mathcal{N}(\textbf{0}, I_{1000}),\thickspace \thickspace t = 1, \hdots, 120, \\
& \negthickspace \negthickspace \mathcal{N}(\textbf{0}, \Sigma), \thickspace \thickspace t = 121, \hdots, 200,
\end{cases} $$

\noindent where $\Sigma$ is a diagonal matrix with  $\Sigma_{kk} = \delta$ if $h_k \in H_D$ and $1$ otherwise, with $\delta = 1.175^2$.

In the third scenario, we generate correlated multivariate Gaussian data with mean change, such that dimensions of $\textbf{y}$ are spatially correlated with one another prior to and after the change: 

 $$ y_t \sim
\begin{cases} 
& \negthickspace \negthickspace \mathcal{N}(\textbf{0}, \Sigma),\thickspace \thickspace t = 1, \hdots, 120, \\
& \negthickspace \negthickspace \mathcal{N}(\mu \textbf{1}_{H_D}, \Sigma), \thickspace \thickspace t = 121, \hdots, 200,
\end{cases} $$

\noindent where we set $\Sigma_{ij} = 0.6^{|i-j|}$ and $\textbf{1}_{H_D}$ to be a $1000 \times 1$ vector whose $k^{th}$ entry is $1$ if $h_k \in H_D $, and $0$ otherwise, and $\mu = \sqrt{\frac{4}{D}}$.  In this case, the total magnitude of mean change in $L_2$ is $\|\mu\textbf{1}_{H_D}\| = 4$.

\begin{figure}[t]
\centering
\includegraphics[scale = 0.205]{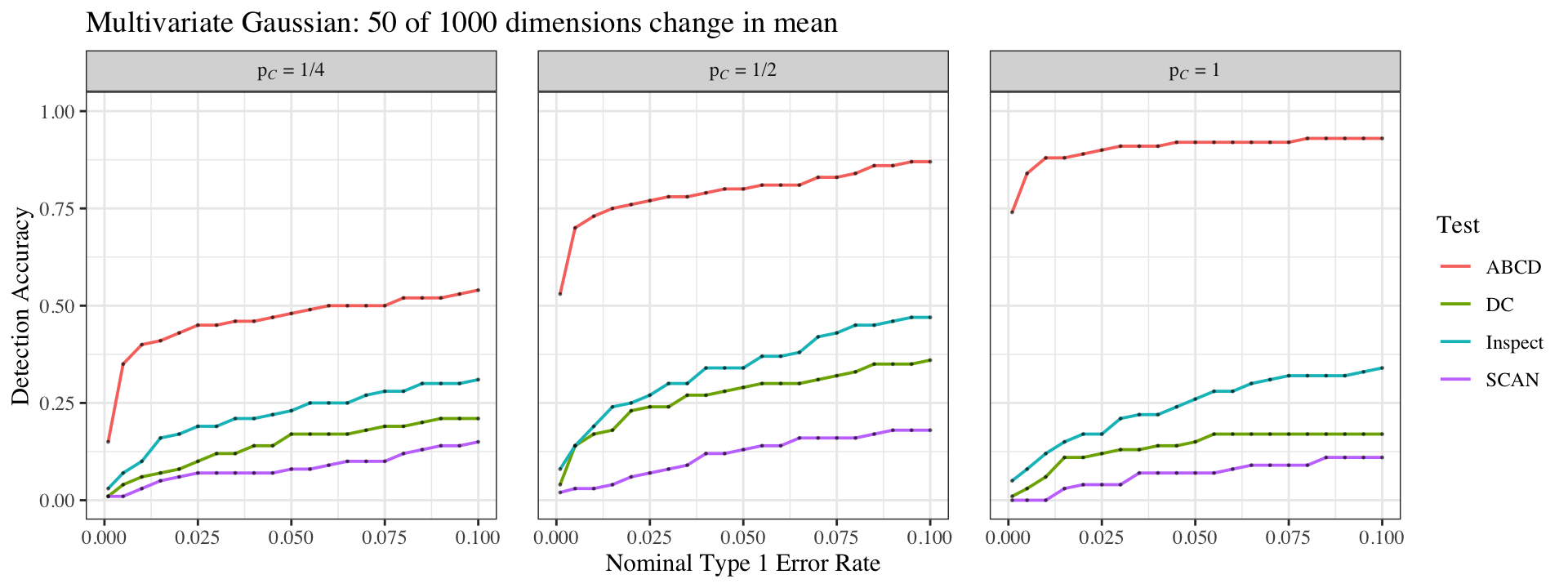}

\vspace{0.25cm}
\includegraphics[scale = 0.205]{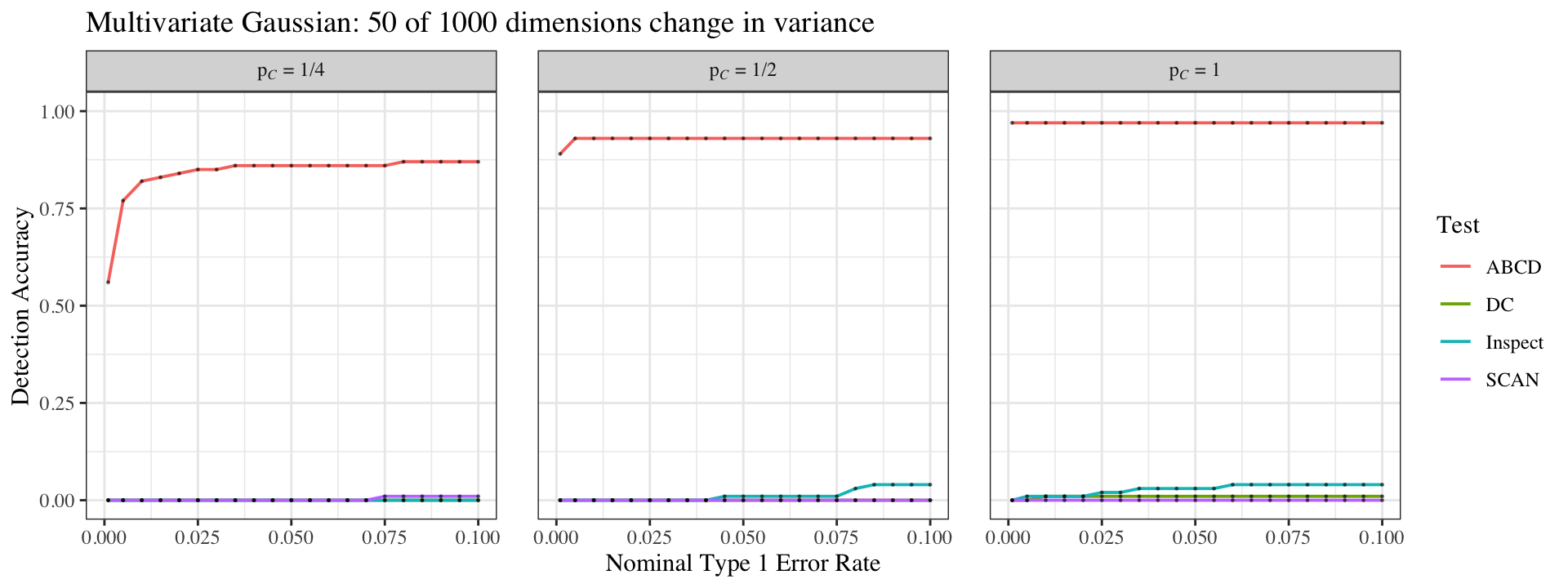}

\vspace{0.25cm}
\includegraphics[scale = 0.205]{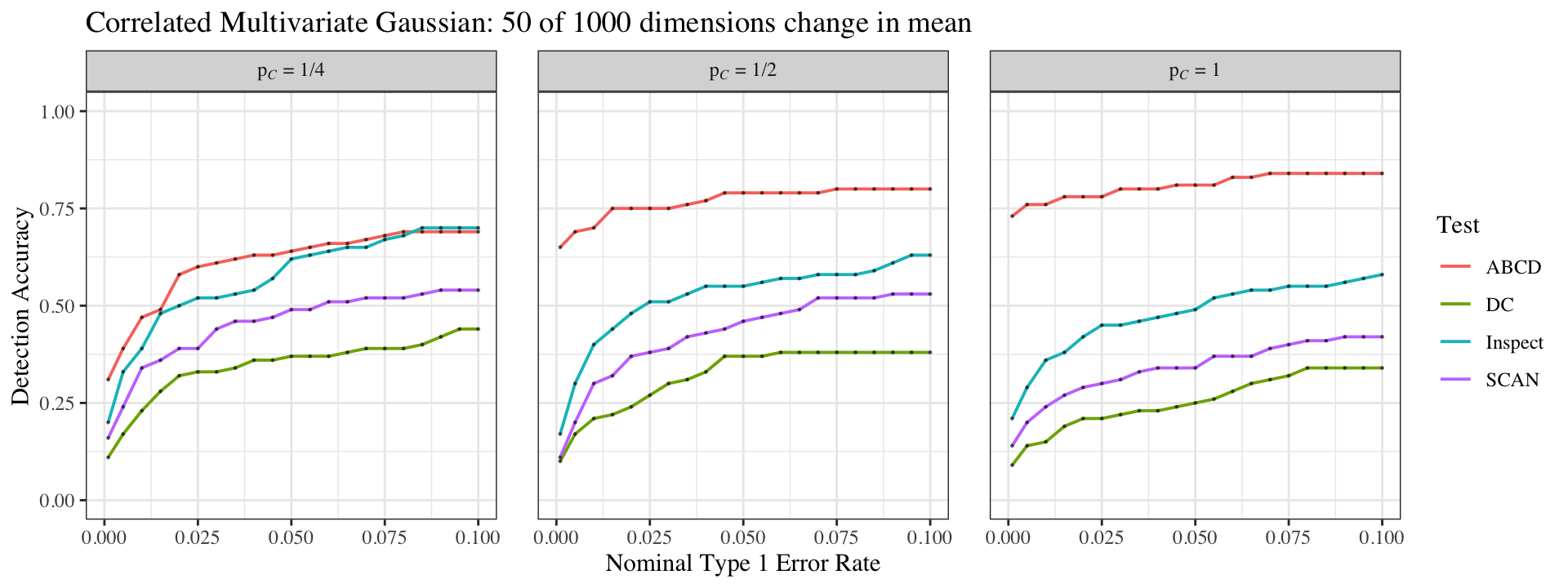}
\caption{Simulation results for multivariate Gaussian data with mean change, multivariate Gaussian data with variance change, and correlated multivariate Gaussian data when $D = 50$. 
We report the proportion of significant change-points within $10$ time-points of $\tau = 120$ for nominal Type-I error rates $0.001 \le \alpha \le 0.1$.}
\label{fig:sim_plots}
\end{figure}

In our single change-point simulations, we compare ABCD with Inspect, DC and SCAN, all recent methods that can detect sparse, spatially clustered change-points in high dimensions. For ABCD, 
we use a $40$-MST and blocking structure vector $\textbf{P} = (1, 4, 10, 20, 40)^T$, to capture a wide range of potential change regions. DC was implemented using the R package \texttt{factorcpt} using their default settings. The Inspect method was implemented using the \texttt{InspectChangepoint} R package, also using their default settings.

Since we know the underlying data distributions, we can directly obtain thresholds for Type-I error control under the null distribution. We generate $10,000$ Monte Carlo simulations for each scenario to obtain thresholds with nominal Type-I error rates between 
$0.001 \le \alpha \le 0.1$ for each method. In practice, if the underlying data distribution is unknown, the null distribution can be approximated using the permutation distribution. 
For each setting, $100$ trials were run. For each method, we report its detection accuracy, which is defined as the proportion of trials that detected a significant change-point at a given level $\alpha$ \textit{and}  estimated $\hat{\tau}$ to occur within $10$ time-points of the true change.

Figure \ref{fig:sim_plots} shows the results for the simulations when $D = 50$. In each scenario, we observe that ABCD's ability to accurately detect significant change-points increases as $p_\mathcal{C}$ increases; this is expected since an increase in $p_\mathcal{C}$ directly corresponds to a more spatially compact change within $\mathcal{C}$. 
The range of blocking structures allows ABCD to capture most of $\mathcal{C}$ within a single block; this results in a substantially higher signal-to-noise ratio within that block, which improves ABCD's ability to successfully detect and estimate a significant change.

In the mean change scenario (first row of Figure \ref{fig:sim_plots}), ABCD outperforms competing methods, even for the challenging case when $p_\mathcal{C} = \frac{1}{4}$. ABCD was also unmatched in its ability to detect changes in variance, as the competing methods are designed to detect mean changes.  When the dimensions under the null are correlated, the performance of ABCD is more comparable to that of competing algorithms. 
However, as $p_\mathcal{C}$ increases, we see that ABCD's detection accuracy begins to outperform existing methods; competing methods actually fare worse for larger values of $p_{\mathcal{C}}$, perhaps due to the resulting increase in spatial correlation between changed components as discussed previously. ABCD counteracts this by isolating the change-point signal and maintains detection ability. 

Additional simulation studies can be found for different levels of sparsity ($D=25$ and $D=100$) and for non-Gaussian data in Supplement Section A3.  
For non-Gaussian data, simulations studies include data generated from multivariate log-normal and multivariate $t$ with $3$ degrees of freedom. We observe that ABCD especially dominates in the sparsest setting, when $D = 25$. The plots exhibit similar trends, in that when the alternative is more spatially compact ($p_\mathcal{C}$ increases), ABCD outperforms competing methods in detection accuracy and power. Additionally, ABCD performs well even in the presence of non-normal data, whereas SCAN and Inspect can experience limited power due to their normality assumptions.

\subsection{Image Simulation}
\label{sec:sim_studies2}

\begin{figure}[h]
\centering
\makebox[\textwidth][c]{\includegraphics[width=0.80\textwidth]{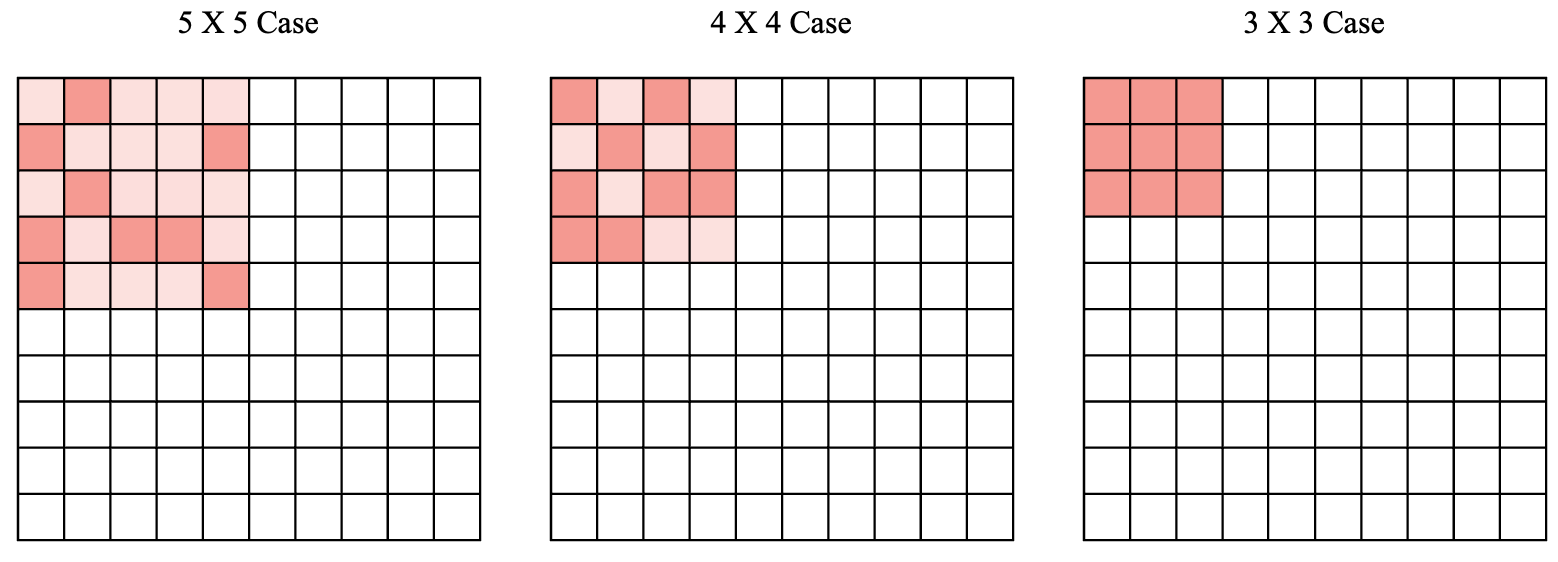}}%
\caption{An illustration showing the change region (in light red) and an example of selected pixels to change (i.e. included in $H_d$) are shown in darker red.}
\label{fig:image_dims_diagram}
\end{figure}

We simulate imaging data and compare ABCD to change-point methods which are able to incorporate spatial information; we consider both the 2WayMOSUM method from \cite{li_2_2024} as well as the SSS method from \cite{kulldorff_scan_2009}. Since the competing methods can be computationally expensive for large time series, we conduct a smaller simulation where in each trial $\textbf{y}$ consists of $n = 200$ observations such that each observation is a $10 \times 10$ image.  A change in mean takes place after $\tau = 120$.  We simulate $\mathbf{y}$ from three distributions:   multivariate standard Gaussian, correlated multivariate Gaussian, and multivariate $t$ with 5 degrees of freedom. As in the example from Section \ref{sec:2d_ABCD_desc} we label these pixels as $(h_{1,1}, h_{1,2},\ldots, h_{10,9}, h_{10,10})$, with the location of the pixel $h_{i,j}$ in the $i^{th}$ row and $j^{th}$ column of the image series taken to be $(i,j)$. 

For each trial, we set the change region, $\mathcal{C}$, to be
a square in the top-left corner of the image. 
We consider three cases for each distribution: $\mathcal{C}$ is a $5 \times 5$, $4 \times 4$ and $3 \times 3$ subset of pixels, as shown in Figure \ref{fig:image_dims_diagram}. In each case we randomly select $D = 9$ dimensions inside the square to change in mean, and denote the index set with $H_D$ once again. 

For the case of multivariate  Gaussian data undergoing mean change, data from pixels in $H_D$ are generated from standard $\mathcal{N}(0,1)$ Gaussian for observations $t = 1,\ldots 120$ and $\mathcal{N}(\mu,1)$ for observations $t = 121, \ldots 200$.  Observations on pixels not in $H_d$ (i.e. unchanged pixels) are $\mathcal{N}(0,1)$ for $t = 1, ..., 200$.  In the scenario of correlated multivariate Gaussian data undergoing mean change, the pixels in $H_d$ similarly have a marginal mean of $0$ for $t = 1, \ldots 120$ and mean of $\mu$ for $t = 121, \ldots 200$, with pixels outside $H_d$ having a mean $0$ for all observations. We specify the variance of each pixel observation to be $1$, with the covariance between any two pixel observations for a given $t$ determined by their Euclidean distance. A given pair of pixels $h_{i_1, j_1}$ and $h_{i_2,j_2}$ with respective locations $(i_1, j_1)$ and $(i_2, j_2)$ have a Euclidean distance $d(h_{i_1, j_1}, h_{i_2,j_2}) = \sqrt{(i_1 - i_2)^2 + (j_1 - j_2)^2}$. We set the covariance between their respective observations at time $t$ to be $0.6^{d(h_{i_1, j_1}, h_{i_2,j_2})}$. Finally for the multivariate $t_5$ distribution data, pixels in $H_D$ are generated from a central $t$ distribution with $5$ degrees of freedom for $t = 1,\ldots 120$ and central $t_5$ with a level shift $\mu$ for $t = 121, \ldots 200$. Pixels outside $H_D$ are generated from a central $t_5$ distribution for $t = 1, ..., 200$. In order to directly compare results across settings, we have set $\mu = \sqrt{\frac{2}{9}}$ in all cases, such that the overall magnitude of mean change in $L_2$ for each image series is $2$.

We compare each method's detection accuracy at level $\alpha = 0.05$ and report the proportion of significant and accurate change-points detected by each method out of $100$ experiments for each setting. As before, we consider a change-point estimate $\hat{\tau}$ to be accurate if $|\hat{\tau} - \tau| \le 10$, with $\tau = 120$. Here ABCD is implemented with a permutation p-value where $U = 1000$, using a $k$-MST with $k=40$ and with blocking structures given by 

$$\textbf{P} =  \begin{bmatrix}
   1 & 2 & 3  \\
   1 & 2 & 3 \\
  \end{bmatrix}^T.
$$

2WayMOSUM was implemented using the $\texttt{L2HDChange}$ package using standard settings for the size of the sliding window and for generating a critical threshold (using $999$ Monte Carlo permutation trials). In order to have fair comparison, 2WayMOSUM is given a list of pixel groups to test that are equivalent to the blocking structures of the ABCD approach. The SSS method was implemented using the \texttt{RSatscan} package and the corresponding \texttt{SaTScan} software. We use its default settings for detection of space-time clusters, with a cylindrical scan window, a Gaussian likelihood ratio test, and Monte Carlo critical threshold generation using $999$ permutation trials. To compare detection ability in the single change-point scenario, we use information from the single most significant cluster detected by SSS in each trial. As this method was not specifically built for change-point detection, it does not provide a single change-point estimate, but instead provides estimated start and end times of a space-time cluster for a subset of pixels, which can be denoted as $(\hat{c}_{start}, \hat{c}_{end})$. This being the case, we set $\hat{\tau} = \hat{c}_{start} - 1$ to assess the accuracy of SSS, as the method should ideally detect a cluster of changed observations starting at time $\tau + 1 = 121$ ending at time $200$. 

 \begin{table}[h]  \centering
 \caption{Number of trials out of $100$ where ABCD, 2WayMOSUM and SSS found a significant and accurate clustered mean change of overall $L_2$ magnitude $2$, with significance level $\alpha = 0.05$, for each setting. Here, $\tau = 120$ and accurate trials are defined as $|\hat{\tau} - 120| \le 10$. }
\begin{tabular}{ |c|c|c|c|c|c|c|c|c|c| } 
\hline
 & \multicolumn{3}{c|}{Gaussian} & \multicolumn{3}{c|}{Correlated Gaussian} & \multicolumn{3}{c|}{$t_5$}  \\ 
\hline
$p_\mathcal{C}$ & $9/25$ & $9/16$ &  $1$ & $9/25$ & $9/16$ &  $1$ & $9/25$ & $9/16$ &  $1$\\ 
\hline
ABCD & \textbf{96} & 95 & \textbf{98} & \textbf{53} & \textbf{65} & \textbf{61} & \textbf{77} & \textbf{81} & \textbf{92} \\ 
\hline
2WayMOSUM & 63 & 57 & 66 & 26 & 41 & 40 & 37 & 37 & 44 \\
\hline
SSS & 92 & \textbf{98} & \textbf{98} & 19 & 34 & 42 & 4 & 21 & 59\\
\hline
\end{tabular}
\label{table:image_comp_full}
\end{table}

The results of this simulation are shown in Table \ref{table:image_comp_full}. We observe that ABCD has superior ability to accurately detect spatially clustered change-points in this single change-point setting relative to 2WayMOSUM in all settings. This is somewhat expected, as in \cite{li_2_2024} the authors acknowledge that their method is developed specifically for multiple change-point detection, and the sliding window approach 2WayMOSUM leads to a loss in power in the single change-point case. 
SSS on the other hand appears to perform at least as well in the multivariate Gaussian setting due to its normality assumption, having a similar proportion of significant and accurate changes detected compared with ABCD. 
On the other hand, in the presence  of non-normal or correlated data, as in the simulations with $t_5$ and correlated Gaussian data, ABCD appears to have the edge over SSS, especially if the changed pixels are not tightly packed in space, such as if $p_\mathcal{C} < 1$. This advantage becomes more apparent as we shift away from normality, in the setting when the pixels are generated from $t$ with 5 degrees of freedom.   

Beyond this, the practical restrictions of the SSS method may pose problems in certain change-point scenarios. For one, pre-processing may be needed for SSS to function properly if the dimensions of the series were heterogenously distributed under $H_0$, for instance if the components had different initial marginal means. Additionally, a requirement of the algorithm as implemented in the \texttt{SaTScan} software is that the detected space-time cluster can span at most $50$ percent of the observations, which may hinder the algorithm's detection accuracy if, for example, a single change occurred near the start of a time series.

\section{Remote Sensing Application}\label{sec:data_app_section}

In September 2020, Iran announced its decision to rebuild a centrifuge assembly center underground at Natanz, following the destruction of the previous facility in  July of that year \citep{albright_imagery_2022}. By analyzing the buildings and activities above this structure, we can gain insights into the likely developments taking place underground, providing valuable information on Iran's progress in its nuclear weapons program at this site. Using Google Earth Engine, we obtained three Sentinel-2 satellite bands covering a section of the Natanz nuclear facility, where an underground centrifuge assembly plant has been under construction in recent years. In particular, we aim to use the Sentinel-2 satellite images to determine exact dates and locations of new construction activity at this site over the period December 2019 to July 2024, as high-resolution reference images from a variety of sources which clearly show construction changes are relatively infrequent during this period.

This objective is substantially challenging for multiple reasons. First, the raw data is immense, with three bands, each containing $112 \times 211$ pixels per observation, collected over nearly five years.  Furthermore, these bands exhibit significantly non-stationarity over time and contain outlier images caused by cloud cover. Ultimately, we want to compile a \textit{single} time series of images which can be analyzed with change-point methods. Moreover, changes of interest are often extremely sparse and occur in an unknown change regions. These challenges necessitate extensive data pre-processing and require a flexible yet powerful, approach that does not rely on strict distributional assumptions or prior knowledge of the change regions. Due to the size and complexity of the data, computational costs and scalability must also be considered.

In Section \ref{sec:natanz_setup}, we outline the steps to pre-process the raw Sentinel-2 data for analysis. Details are provided on how to integrate ABCD with seeded binary segmentation to detect multiple change-points \citep{kovacs_seeded_2023}. In Section \ref{sec:data_app_results}, we examine the change-point detection results in the Natanz remote sensing application for ABCD and other competing algorithms. 

\subsection{Pre-processing and Analysis}\label{sec:natanz_setup}

Significant pre-processing of the data was necessary to minimize the effect of outliers and protect against non-stationarity. To minimize outliers, we filter out observations where significant cloud cover was present using functions from Google Earth Engine. Due to this filtering and the irregular Sentinel-2 recording frequencies, the time between observations in our data sequence is irregularly spaced. However, there are no gaps of several months between images, allowing us to maintain reasonably accurate estimates of when buildings, roads, or other structures are newly constructed. After filtering out particularly cloudy observations, we have $n = 335$ image observations, each with $d_1 = 112$,  $d_2 = 211$ for each band.  We focused on the Sentinel-2 bands $B2$, $B3$ and $B4$, which represent the natural color spectrum. Thus, the raw data can be represented as the four-dimensional array $A_{3 \times 112 \times 211 \times 335}$.

The raw Sentinel-2 bands are highly non-stationary primarily for two reasons. First, there is a notable jump in pixel values across all bands around January 2022, caused by an alteration in how Sentinel-2 records its band data. Second, seasonal patterns in the mean and variance of pixel values, likely driven by temperature fluctuations on the ground, also contribute to non-stationarity. To address this,
we utilize a robust method to standardize the images and reduce non-stationarity. We define a new array $\tilde{A}_{3 \times 112 \times 211 \times 335}$ of equivalent size, which represents the standardized version of $A$. For an image $A_{i,\cdot,\cdot,t}$ from array $A$ for a given band $i$ at time $t$, we can define each pixel in $\tilde{A}_{i,\cdot,\cdot,t}$, indexed by $j = (1, ..., 112)$ and $k = (1, ..., 211)$, as 

$$\tilde{A}_{i,j,k,t} \equiv \frac{A_{i,j,k,t} - \text{median}(A_{i,\cdot,\cdot,t})} {\hat{F}^{-1}_{0.95}(A_{i,\cdot,\cdot,t}) - \hat{F}^{-1}_{0.05}(A_{i,\cdot,\cdot,t})},$$

\noindent where $\hat{F}$ denotes the ECDF; the empirical 95\% and 5\% quantiles are denoted as $\hat{F}^{-1}_{0.95}(A_{i,\cdot,\cdot,t})$ and $\hat{F}^{-1}_{0.05}$, respectively. 
We implement this standardization for each band $i \in (1,2, 3)$ at each time-point $1 \leq t \leq 335$. 

Next, we create another three-dimensional array denoted by $L_{112 \times 211 \times 335}$,  where for each pixel and time-point 

$$L_{j,k,t} = 
\begin{cases} 
& \negthickspace \negthickspace 1\thickspace \thickspace \text{if construction is known to exist at pixel $(j,k)$ at time t}, \\
& \negthickspace \negthickspace 0 \thickspace \thickspace \text{otherwise}  .
\end{cases}$$ 

This was done by manually classifying areas of construction using high-resolution reference images compiled in Google Earth of the Natanz site taken on 12/18/2019, 12/29/2020, 01/25/2022, 12/22/2022 and 09/16/2023. This array is essentially a rough approximation of the ground truth for where construction was present over time. It is important to note that if construction occurred at a given pixel after a reference image was taken, this pixel only takes a value of 1 once the \textit{next} reference image is taken, when we have visual confirmation that a construction had taken place. 

We then fit a logistic regression model for each pixel in the label array over all time-points, using the three bands of pixel values as covariates, to estimate the probability that a construction is present at a given pixel and time-point. We store the fitted values of these regression models in the three-dimensional array $\hat{L}_{112\times 211 \times 335}$.  Explicitly, for a given pixel $(j, k)$, $j \in (1, ..., 112)$, $k \in (1, ..., 211)$, we fit the model

$$\hat{L}_{j,k,\cdot} = \frac{\exp(a^T \beta)}{1 + \exp(a^T \beta)},$$

\noindent with $a^T = (1, \tilde{A}_{1,j,k,\cdot},\tilde{A}_{2,j,k,\cdot},\tilde{A}_{3,j,k,\cdot})$ and $\beta = (\beta_0, \beta_1, \beta_2, \beta_3)$, $\beta \in \mathbb{R}^4$.  The array $\hat{L}_{112\times 211 \times 335}$ is used as the time series for our final change-point analysis.

To implement ABCD and competing methods for multiple change-point detection, we employ the recently proposed seeded binary segmentation (SBS) algorithm proposed by \cite{kovacs_seeded_2023}. Traditional binary segmentation, introduced by \cite{scott_cluster_1974}, has long been a established method for to extend single change-point tests to a multiple change-point detection framework. However, its recursive nature can result in biased change-point estimates.  Wild binary segmentation (WBS) proposed by \cite{fryzlewicz_wild_2014} ameliorates this issue by randomly selecting sub-intervals of the whole series, however, it can be computationally expensive to implement.  SBS builds off of WBS but instead tests over a deterministic set of sub-intervals, which are on average of smaller length than in WBS. This results in an unbiased multiple change-point detection approach which is markedly faster from a computational standpoint and, like other binary segmentation algorithms, can be paired with any single change-point detection method.  

We compare DC, Inspect, and gSeg to ABCD for this data application, pairing each of these methods with SBS to detect multiple change-points. We were unable to compare ABCD to methods incorporating spatial information, such as 2WayMOSUM or SSS, as such methods were computationally infeasible for data of this scale. We use the standard settings described in \cite{kovacs_seeded_2023} to generate the deterministic sub-intervals and implement their proposed greedy selection method to estimate multiple change-points.  Letting $n_s$ denote the length of a sub-interval, we restrict our simulation to test only on  intervals where $n_s \geq 30$, to limit the amount of false positives. Both ABCD and gSeg are implemented utilizing a $k$-MST similarity graph and the $M(t)$ graph-based statistic. Here we set $k = \lfloor 0.2n_s \rfloor$ for each sub-interval test. For ABCD, we set

$$\textbf{P} =  
  \left[ {\begin{array}{cccc}
   1 & 4 & 8 & 16 \\
   1 & 6 & 12 & 24\\
  \end{array} } \right]^T,
$$
\noindent outlining 4 blocking structures for detecting different sizes of change regions. DC and Inspect are implemented using standard settings from their respective \texttt{R} packages, and implemented on a version of the image time series with vectorized observations.

\begin{figure}[h]
\begin{table}[H] \centering
\caption{The eight most significant change-points detected by ABCD in chronological order. If changes were  detected (within a margin of two observations) by a competing method, a green check is shown in the corresponding column. Bolded change-points indicate at least one competing method failed to detect the change. Change-points in bolded red were detected only by ABCD. Corroborating evidence for construction occurring at these times is provided via visual analysis of heatmaps and  reference images. The type of development ABCD detected is determined based on reports by \cite{albright_imagery_2022} and \cite{albright_imagery_2024}.}
\centering
\makebox[\textwidth][c]{\includegraphics[width=0.99\textwidth]{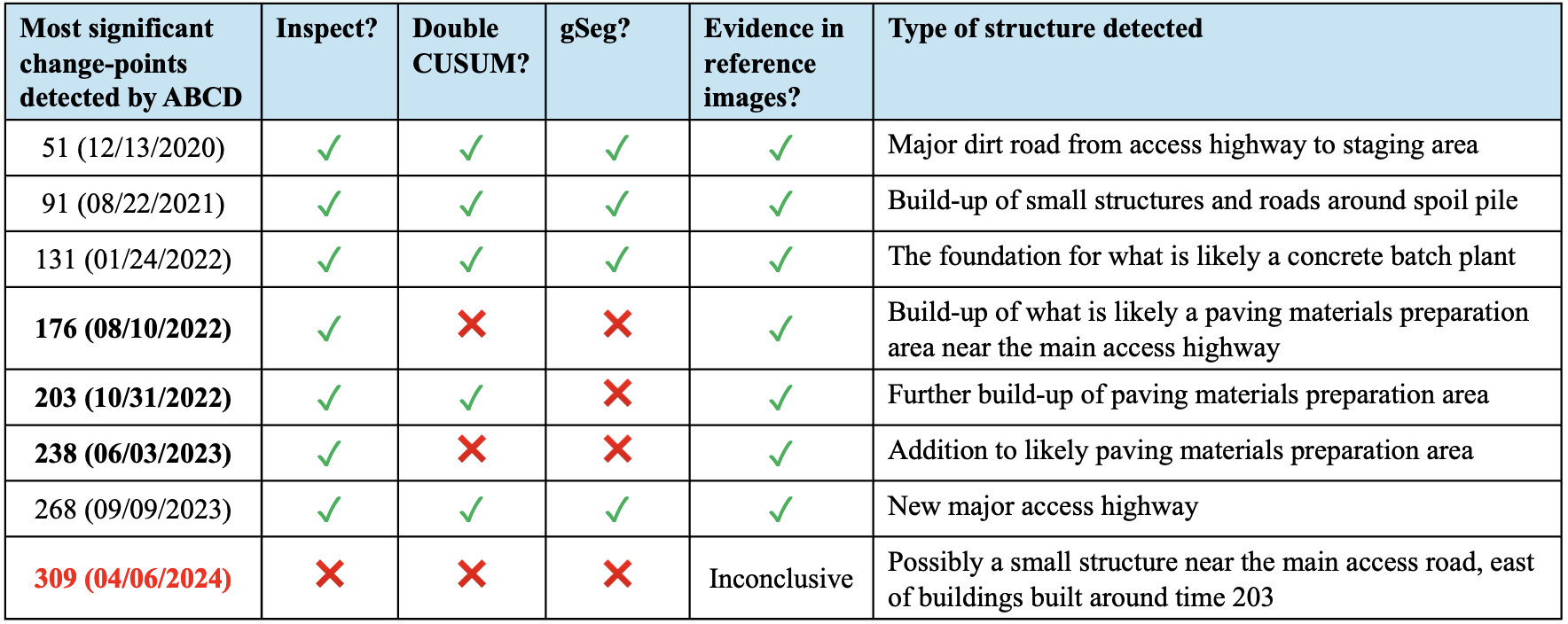}}
\end{table}
\end{figure}

\subsection{Application Results}\label{sec:data_app_results}

\begin{figure}[h]%

    \centering
    \makebox[\textwidth][c]{\includegraphics[width=0.99\textwidth]{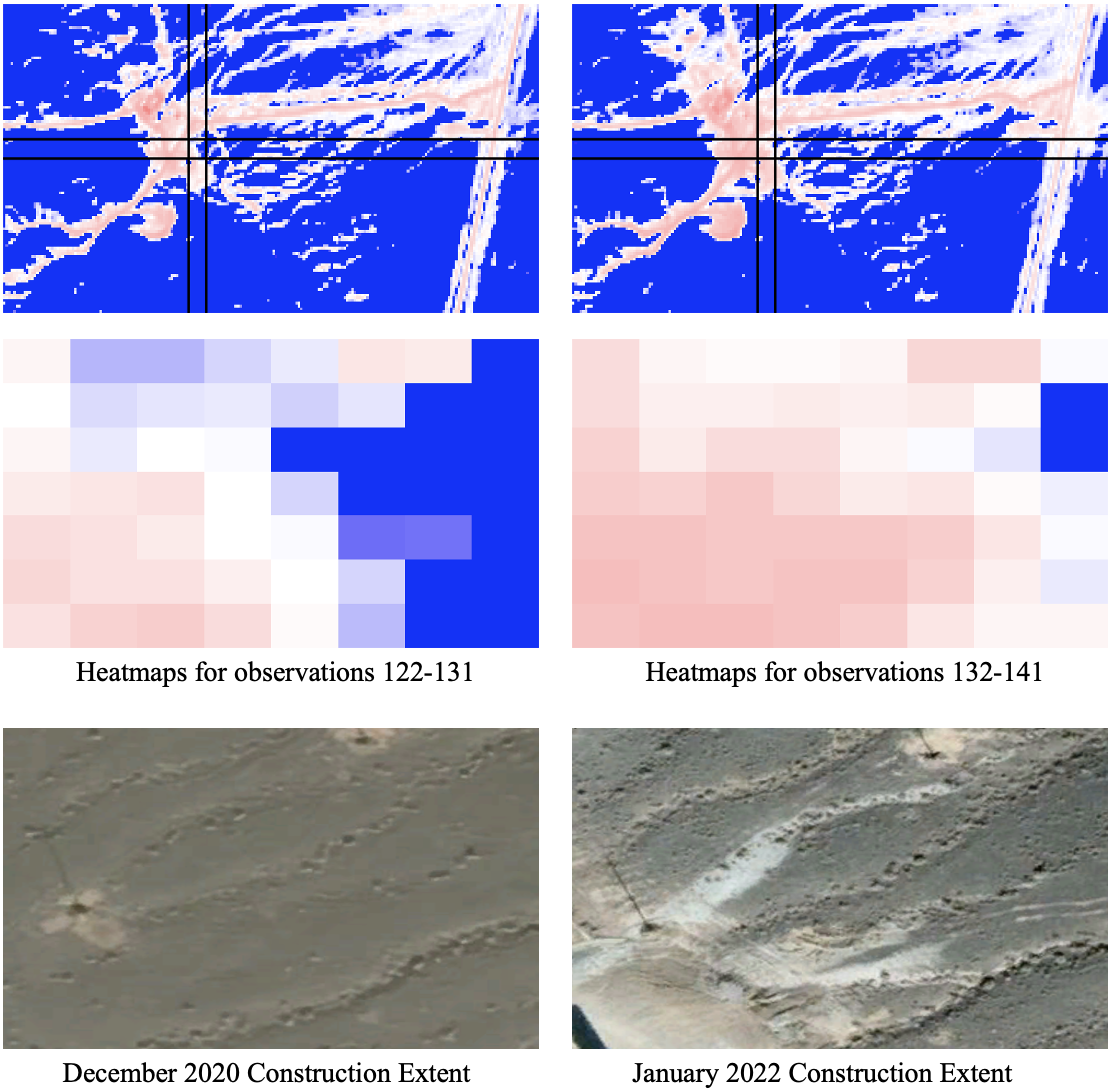}}
\caption{Visualization of the most significant change-point found by ABCD  at time $t = 131$ (01/24/2022).  We identify the block with the maximum graph-based statistic across all blocks. Within this block, we compare heatmaps of the average of ten log $\tilde{W}$ images immediately before and after the change-point. We observe a clear indication of expansion of a newly paved area in the lower-left corner of the block, which we can see has been developed within this time frame using reference images.}
\label{fig:131_natanz}
\end{figure}

Table 3 presents the eight most significant change-points detected by the ABCD algorithm, along with whether competing methods detected the same change-points within two time-points. Since we do not know the underlying distribution, we rely on permutation resampling to obtain thresholds to assess significance; we set $\alpha = 0.01$ and the number of permutations $U = 1000$ for all methods. The majority of change-points identified by ABCD were strongly supported by visual evidence, which suggests they represent valid new developments at Natanz. 

To collect evidence to validate a given change-point detected by ABCD, we first identify the block with the largest graph-based test statistic across all blocking structures, which represents the most likely location of the change. We then visualize this block over time using heatmaps to further confirm whether construction occurred at this location at time $\hat{\tau}$. To achieve this, we define $\tilde{W}_{112 \times 211 \times 335}$ as the mean of pixel values of $\tilde{A}$ values across the three Sentinel-2 bands, for each pixel location and time-point; this is to be used for data visualization. We apply a logarithmic transformation to each pixel value in the array $\tilde{W}$  to generate images at specific time-points that enhance the visibility of hotspots. To prevent undefined values, we set a minimum value for any given pixel in the log-transformed images. Finally, we visually examine heatmaps of these images before and after the estimated change-point within the suspected block containing new activity. 

\begin{figure}[h]%

    \centering
    \makebox[\textwidth][c]{\includegraphics[width=0.99\textwidth]{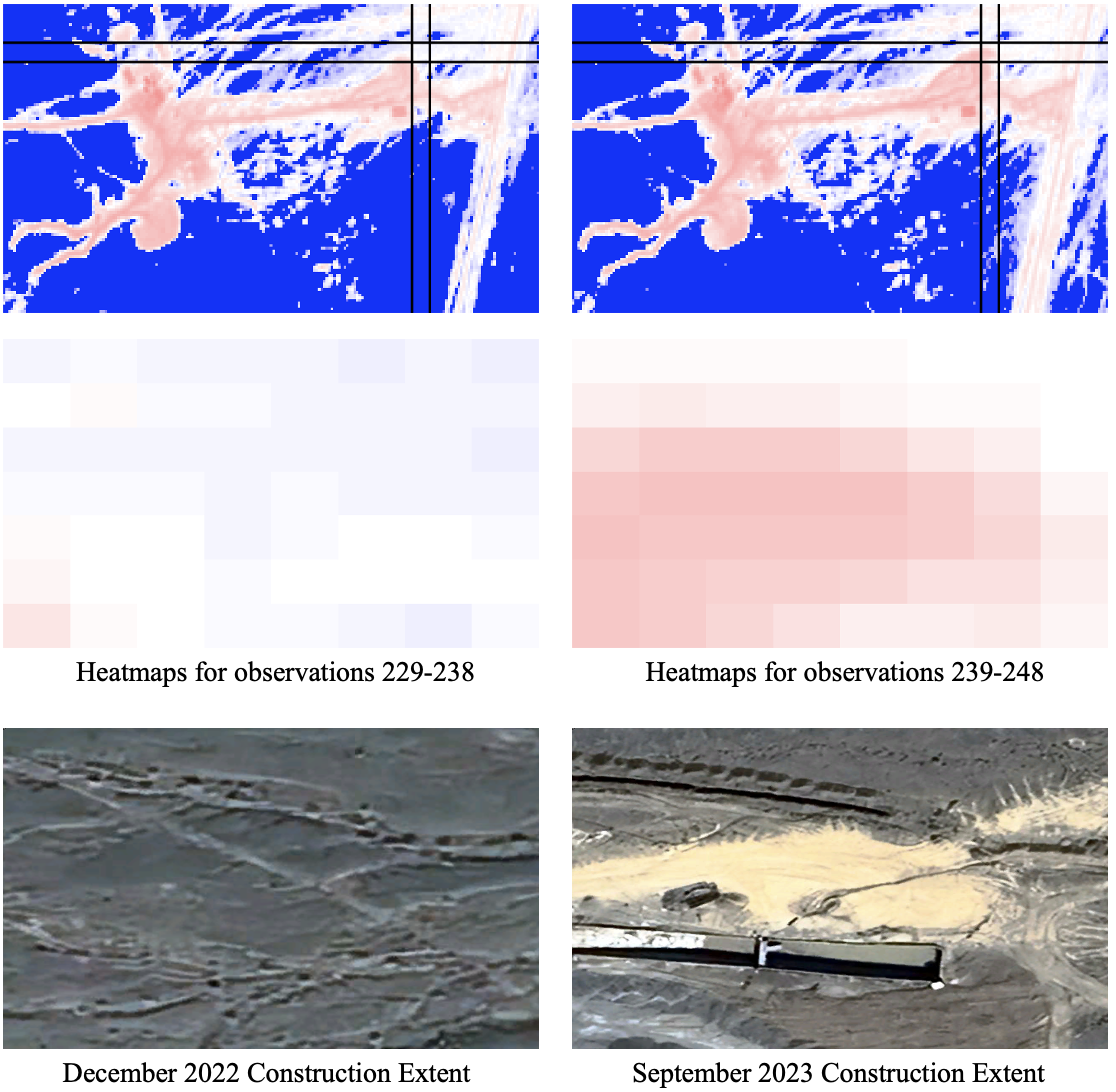}}
\caption{Visualization of a sparse change-point found by ABCD  at  $t = 238$ (06/03/2023).  We identify the block with the maximum graph-based statistic across all blocks at $\hat{\tau} = 238$. Within this block, we compare heatmaps of the average of ten log $\tilde{W}$ image immediately before and after the change-point. Note the outline of a newly paved area within this time frame in the reference images.}
\label{fig:238_natanz}
\end{figure}

We illustrate this procedure in Figure \ref{fig:131_natanz} for the single most significant changepoint at $\hat{\tau} = 131$ (01/24/2022). This was also the most significant change detected by Inspect and was detected in the SBS testing procedure by gSeg and DC. At $\hat{\tau} = 131$ , we identify the block reporting the largest graph-based statistic. To validate this estimated change region, we examine the mean of ten $\log\tilde{W}$ images within this block before and after $\hat{\tau} = 131$. It is clear from Figure \ref{fig:131_natanz}, that there is an outline of a new construction in the lower-left segment of the block. We observe that these developments can also be visualized in high-resolution reference images; this detected change is likely the foundation of a concrete batch plant as described in a report by \cite{albright_imagery_2022}. This demonstrates ABCD's distinctive capability to not only detect change-points, but also to provide estimates of their spatial locations.

The results demonstrate that ABCD can detect highly sparse change-points, such as the estimated change-point at $\hat{\tau} = 238$ (06/03/2023). 
Again, the block reporting the largest graph-based statistic at $t = 238$ out of all blocks is obtained. In Figure \ref{fig:238_natanz}, we compare the mean of ten $\log\tilde{W}$ images within this block before $t = 238$ to the mean of ten $\log \tilde{W}$ images after this time-point; we see that the outline of a new construction on the left side of this block is apparent. Furthermore, we are able to visualize that a new paved area was constructed some time between December 2022 and September 2023 with reference images, providing supportive evidence of the validity of change-point detected by ABCD. 

Crucially, while change-points that competing methods found often agreed with ABCD's results, determining if these change-points were representative of real constructions was much more difficult compared with ABCD, since these methods were unable to provide direct information on the spatial sub-region of the image where the change occurred. ABCD's unique ability to provide locational estimates of where these changes likely occurred makes it a practical tool for spatial change-point analysis. Altogether, this data application illustrates that ABCD performs at least as well as state-of-the-art sparse change-point methods, such as Inspect and DC, while also providing additional change-points undetected by competing methods. 

There were many additional significant change-points detected by ABCD. An analysis of the full results of this data application study, including heatmaps and Google Earth reference images for each detected change-point, can be found in Section A4 of the Supplement.

\section{Conclusion}\label{sec:Discussion}

Motivated by the need to detect small signals in remote sensing imagery, we introduce ABCD as a flexible and powerful non-parametric method for detecting and estimating spatially clustered changes in a time series of data objects. Our work ensembles graph-based scan statistics via an adaptive blocking procedure to target potentially sparse and spatially clustered changes. We do not resort to an exhaustive testing procedure, but instead develop an approach to aggregate signal information across different blocking structures, enabling our method to efficiently capture change regions of unknown size and location. To the best of our knowledge, our method is the only one designed  to detect $\textit{general}$ distributional changes when targeting spatially clustered change-points with no distributional assumptions, making it broadly applicable and effective in a wide range of applications. Our approach has been shown to effective in a various simulation studies compared to other state-of-the art change-point methodology. 

While ABCD has been tailored to the offline setting, there may be significant interest in an online version of such a procedure, especially in the context of satellite image monitoring. In such a setting, ongoing detection of activities or constructions could take place as new satellite images are collected, allowing for `real-time' feedback on observed anomalies. We reserve this is as a future line of research for further exploration. 

\bibliographystyle{apalike} 
\bibliography{cp}  

\end{document}